\documentclass[fleqn,10pt]{wlscirep}
\usepackage[utf8]{inputenc}
\usepackage[T1]{fontenc}

\usepackage[utf8]{inputenc}

\usepackage{mathtools}

\usepackage{booktabs} 
\usepackage[rightcaption]{sidecap}
\usepackage{bold-extra}
\usepackage{amsmath,amsfonts,amssymb}

\usepackage{multirow}
\usepackage{xcolor}
\usepackage{xspace}

\usepackage{graphicx}
\usepackage{grffile}
\usepackage{siunitx}

\usepackage{epstopdf}

\usepackage{boxedminipage}

\usepackage{tabularx}
\usepackage{array}
\usepackage{tabularx}
\usepackage{array}

\newcolumntype{P}[1]{>{\centering\arraybackslash}p{#1}}
\newcolumntype{M}[1]{>{\centering\arraybackslash}m{#1}}
\usepackage{hyperref}
\usepackage{textcase}
\usepackage{caption}

\usepackage{stfloats}
\usepackage[noabbrev, capitalise]{cleveref}

\usepackage{nccmath}
\usepackage{subcaption}
\usepackage{wrapfig}
\usepackage{lipsum}

\usepackage{soul}
\definecolor{shadecolor}{rgb}{0.7421875,0.7421875,0.7421875}

\title{Temporal Motifs in Patent Opposition and Collaboration Networks}

\author[1]{Penghang Liu}
\author[2, 3, 4]{Naoki Masuda}
\author[4]{Tomomi Kito}
\author[1*]{Ahmet~Erdem~Sar{\i}y\"{u}ce}
\affil[1]{Department of Computer Science and Engineering, State University of New York at Buffalo, Buffalo, NY, USA}
\affil[2]{Department of Mathematics, State University of New York at Buffalo, Buffalo, NY, USA}
\affil[3]{Computational and Data-Enabled Science and Engineering Program, State University of New York at Buffalo, Buffalo, NY, USA}
\affil[4]{Faculty of Science and Engineering, Waseda University, Tokyo, Japan}

\affil[*]{erdem@buffalo.edu}

\begin{abstract}
Patents are intellectual properties that reflect innovative activities of companies and organizations. 
The literature is rich with the studies that analyze the citations among the patents and the collaboration relations among companies that own the patents.
However, the adversarial relations between the patent owners are not as well investigated.
One proxy to model such relations is the patent opposition, which is a legal activity in which a company challenges the validity of a patent.
Characterizing the patent oppositions, collaborations, and the interplay between them can help better understand the companies' business strategies.
Temporality matters in this context as the order and frequency of oppositions and collaborations characterize their interplay.
In this study, we construct a two-layer temporal network to model the patent oppositions and collaborations among the companies.
We utilize temporal motifs to analyze the oppositions and collaborations from structural and temporal perspectives.
We first characterize the frequent motifs in patent oppositions and investigate how often the companies of different sizes attack other companies.
We show that large companies tend to engage in opposition with multiple companies.
Then we analyze the temporal interplay between collaborations and oppositions.
We find that two adversarial companies are more likely to collaborate in the future than two collaborating companies oppose each other in the future.
\end{abstract}
\begin{document}

\flushbottom
\maketitle

\thispagestyle{empty}

\section*{Introduction}
Innovative activities of companies and organizations are often reflected by patents.
Once a patent is granted, the owner receives a temporary right to legally exclude others from making, using, or selling an invention.
Therefore, managing patents is a crucial aspect in enhancing the competitiveness of companies.
Thanks to the recent advances in computer and data science, large-scale patent data have been analyzed with the aim to understand innovation processes~\cite{youn2015invention, hasan2010innovation}. 

Many studies have investigated the network of patents, in order to understand the evolution of technology and assess the values of inventions~\cite{gao2018community, verspagen2007mapping, funk2017dynamic, erdi2013prediction, acemoglu2016innovation}.
In a patent citation network, patents are connected via citations, and the technological importance of a patent is often measured by its number of citations received.
Patents can also reflect the relations between the patent owners.
Co-ownership of patents by multiple companies implies collaborative relations among them.
Considerable efforts have been made to analyze the patent collaboration networks, in which the edges are positive and symmetric relationships rather than citation relationships, and discover the factors that lead to collaborations among companies~\cite{Balconi2004, Fleming2007, Graf2011}.
However, the rival or adversarial relations among patent owners have not been addressed as much. 
The adversarial relations among companies can be captured by patent opposition, which is a legal activity in which a company challenges the validity of a patent in usually 6-9 months after the grant date. Since the challenged patent may be revoked if the opposition is successful, companies often oppose the patents owned by rival companies to try to hinder their innovation activities~\cite{helmers2018, Sterlacchini2016, Harhoff2003}.
In a recent study, Kito et al.~\cite{kito2020inter} constructed a network of companies and analyzed adversarial relationships in patent networks. They investigated the global and local statistics of a patent opposition network and applied network motif analysis among other things. However, the interplay between patent oppositions and collaborations, especially the temporal correlation between them, has not thoroughly been studied.

The business relations among companies are highly intricate, where rivalries and cooperations are both common and heavily intertwined with each other.
Our goal is to characterize the patent oppositions, collaborations, and the interplay between them which may help us understand how companies interact with each other and strategize their business decisions.
While there may exist other methods to achieve this, potentially including various social physics approaches~\cite{perc2019social}, network-based approaches have been shown to give promising results~\cite{kito2020inter}.
In this work, we embrace network-based analysis and specialize on temporal aspects of the networks.
Temporal network analysis is expected to be a useful approach in this context for two reasons: (1) the relations among companies often change over time, and it is not possible to understand their dynamic structure via the static network representation; and (2) a company's business strategy is reflected by its reactions to the previous oppositions and collaborations, the understanding of which requires the knowledge of the chronological order and the time difference between oppositions and collaborations.

Network motifs are defined as statistically overrepresented subgraphs~\cite{milo2002network}, which are often used for examining interactions among small sets of nodes in the networks.
Several models have been proposed for motifs in temporal networks~\cite{K11, S14, H15, P17}, which we compared in a recent survey~\cite{liu2021temporal} and were also partially covered in earlier work~\cite{holme2012temporal, holme2015modern}.
Temporal motif analysis has been shown to be a versatile tool for many applications including cattle trade movements~\cite{bajardi2011dynamical}, editor interactions in Wikipedia~\cite{jurgens2012temporal}, mobile communication networks~\cite{kovanen2013temporal, li2014statistically}, and human interactions~\cite{zhang2015human} (In the following text, for the sake of simplicity, we follow the recent literature on temporal motifs~\cite{P17, liu2021temporal} to refer to all temporal network patterns of limited size as motifs, not only the overrepresented patterns as suggested by Milo et al.~\cite{milo2002network}.)

In this study, we construct a two-layer temporal network to model the patent oppositions and collaborations from 1978 to 2018. We deploy temporal motifs to analyze the oppositions and collaborations from structural and temporal perspectives. 
Our analysis is two-fold.
First, we identify frequent temporal motifs in patent oppositions using different timing parameters and then investigate the sizes of the companies to further understand the opposition relations.
Second, we identify the collaborations between the companies in the detected opposition motifs and investigate the interplay between oppositions and collaborations.

\section*{Methods}

\subsection*{Patent opposition and collaboration network}

\begin{figure*}[t!]
(a)
\includegraphics[width=0.65\linewidth]{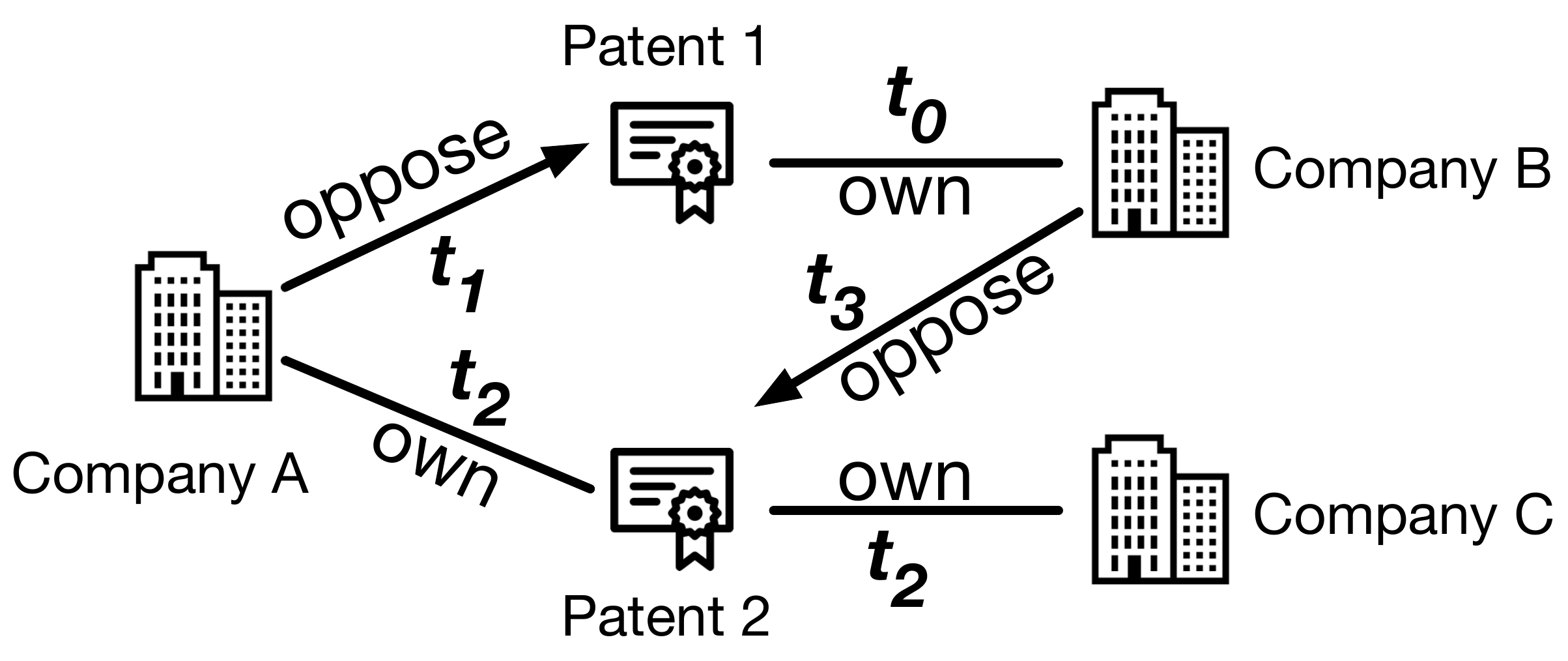}
(b)
\includegraphics[width=0.25\linewidth]{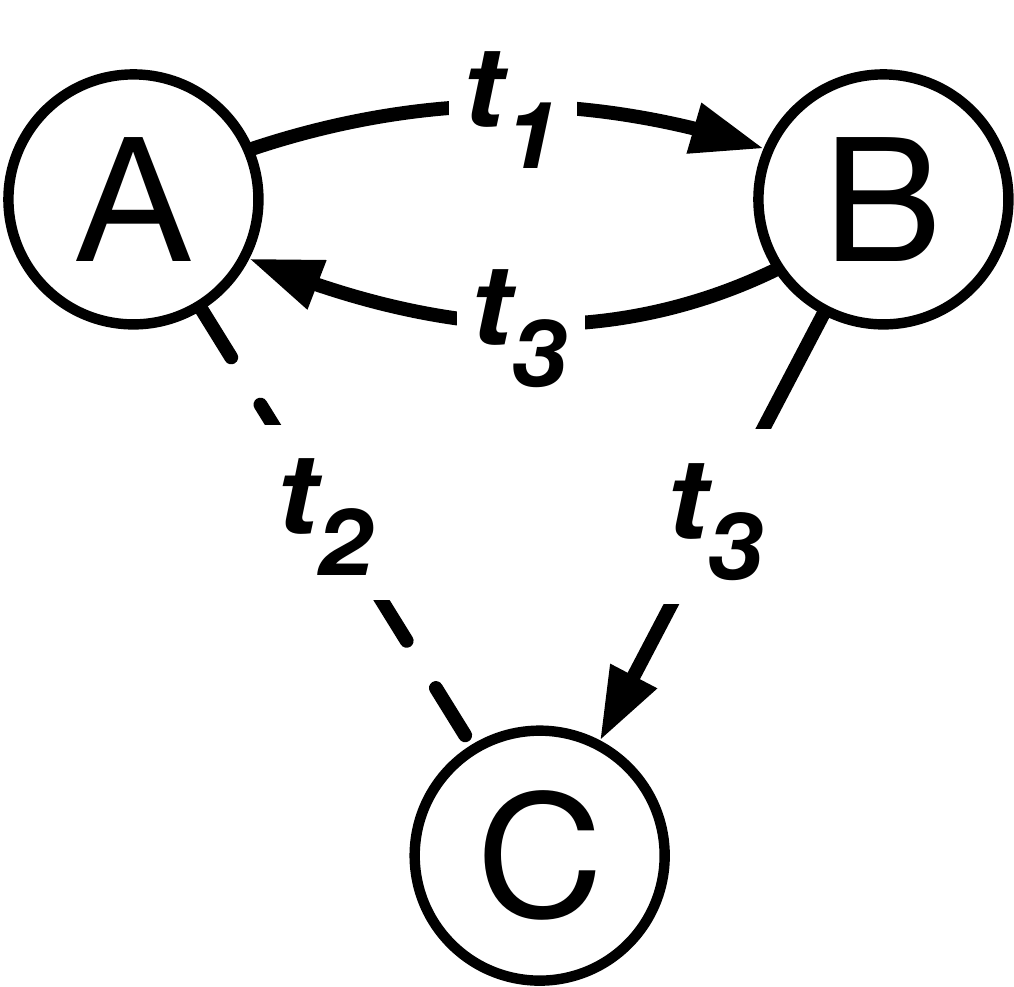}
\caption{\it (a) An example for patent oppositions and collaborations. Company A opposes a patent owned by company B at $t_1$,  and then companies A and C co-own a patent at $t_2$, which is opposed by company B at $t_3$. (b) The corresponding patent network. The directed solid lines represent patent oppositions. The undirected dashed line represents the collaboration.}
\label{fig:toy}
\end{figure*}

We explore temporal networks that we represent as $G = (V, E)$, where $V$ is the set of nodes, and $E$ is the set of time-stamped events.
Each event $e_i \in E$ is a 3-tuple $(u_i, v_i, t_i)$~\cite{Masuda16}], which represents a directed relation from the source node $u_i$ to the target node $v_i$ that occurs at time $t_i$.
The set $E$ is a time-ordered list of $k$ events such that $t_1 \le t_2 \le t_3 \le \cdots \le t_k$.
Here we distinguish edges and events, where the edge $(u, v)$ is the static projection of an event $(u, v, t)$, and there may be multiple events on an edge. 
We restrict ourselves to opposition events in our temporal motif analysis. We consider collaboration events only within the context of temporal opposition motifs.

We represent the two-layer temporal network of patent oppositions and collaborations as $G = (V, E_O, E_C)$, where $V$ is the set of nodes representing the companies. 
The event set $E_O$ defines the opposition layer, in which each event $(u, v, t) \in E_O$ indicates a directed relation that company $u$ opposed a patent owned by company $v$ at time $t$. 
The event set $E_C$ defines the collaboration layer, in which each event $(v_1, v_2, t) \in E_C$ indicates an undirected relation that company $v_1$ and company $v_2$ have co-owned a patent since time $t$.
We consider a patent is owned by a company since time $t$ if either the patent is granted to the company at $t$ or the patent is transferred to the company at $t$.
\cref{fig:toy} shows an example for patent oppositions and collaborations among three companies: company A opposes company B at time $t_1$, company A and C collaborate at time $t_2$, and then company B opposes company A and C at time $t_3$.
Note that the patent oppositions and collaborations can also be represented as a signed network. For example, Park et al. analyze the motifs in a signed directed friendship network~\cite{park2021motif}. However, we represent the patent oppositions and collaborations as a two-layer network rather than a signed network because there may be both opposition and collaboration between the same pair of nodes, which cannot be addressed by a signed network.

We use a previously collected patent data from Orbis Database~\cite{orbis} and Orbis Intellectual Property Database~\cite{orbisip}. We show properities of the two-layer temporal network of patent oppositions and collaborations in~\cref{tab:data}. The data contains 11,317 companies, 26,433 oppositions, and 1,554 collaborations. There are 14,320 and 1,009 edges in the static projection of the opposition and collaboration layers, respectively, which indicates that there are repetitive oppositions and collaborations among the companies.
The low average degree indicates that the two-layer network is sparse. However, there is a giant connected component in the static projection of the opposition network which contains 66\% of nodes and 81\% of edges, and all the other components have less than 10 nodes. Kito et al.~\cite{kito2020inter} show that many companies own and oppose patents in multiple industrial fields. Hence the network of patent oppositions and collaborations is not segregated with respect to technical domains but rather heavily intertwined.

\begin{table*}[!h]
\centering
\begin{tabular}{|c||r|r|r|c|}
\hline
Layer    & \# nodes  & \# edges  & \# events & Time span \\ \hline
Opposition & 11,317 & 14,320 & 26,433 & 2/25/1981 - 4/19/2018 \\ \hline
Collaboration    & 11,317    & 1,009  & 1,554  & 6/23/1978  - 1/10/2018 \\ \hline
\end{tabular}
\caption{\it Properties of the two-layer temporal network of patent oppositions and collaborations.}
\label{tab:data}
\end{table*}

\subsection*{Motifs in temporal networks}
In this study, we consider two types of timing thresholds for temporal motifs: The inter-event time threshold limits the time difference between any pair of consecutive events in the motif, and the time-window threshold requires all the events to occur within a given time interval. Formally, given a temporal network $G = (V, E)$ and two timing thresholds $\Delta_C$ and $\Delta_W$, which we refer to as the inter-event time threshold and time-window threshold, respectively, an $n$-node $m$-event (with $n \geq 2$ and $m \geq 2$) motif $M = (V', E')$, where $E' = \{ (u'_1, v'_1, t'_1), \ldots, (u'_m, v'_m, t'_m) \}$ and $t'_1 \le t'_2 \le \cdots \le t'_m$, is a temporal subgraph of $G$ such that

\begin{itemize}
\item $|V'| = n$, $|E'| = m$, $V' \subseteq V$, and $E' \subseteq E$.
\item For any pair of consecutive events $(u'_i, v'_i, t'_i) \in E'$ and $(u'_{i+1}, v'_{i+1}, t'_{i+1}) \in E'$ such that \{$u'_i, v'_i\} \cap \{u'_{i+1}, v'_{i+1}\} \neq \emptyset$, it holds true that $t'_{i+1} - t'_{i} \leq \Delta_C$.
\item For the first and the last events $(u'_1, v'_1, t'_1) \in E'$ and $(u'_m, v'_m, t'_m) \in E'$, it holds true that $t'_m - t'_1 \leq \Delta_W$. 
\item For any pair of events $(u',v',t') \in E'$ and $(u'', v'', t'') \in E'$, $t' \neq t''$.
\end{itemize}

\noindent The last statement in the case of the opposition network requires all oppositions in the motif to have unique timestamps. This is because if an opposed patent is owned by several companies, an opposition event from the opposing company to each of the owner companies occurs at the same time.
Allowing these opposition events to belong to the same motif is not helpful since they provide redundant information from the same patent opposition.

\begin{figure*}[t!]
\centering
\begin{tabular}[b]{cc}
\begin{tabular}[b]{c}
(a)
\includegraphics[width=0.38\linewidth]{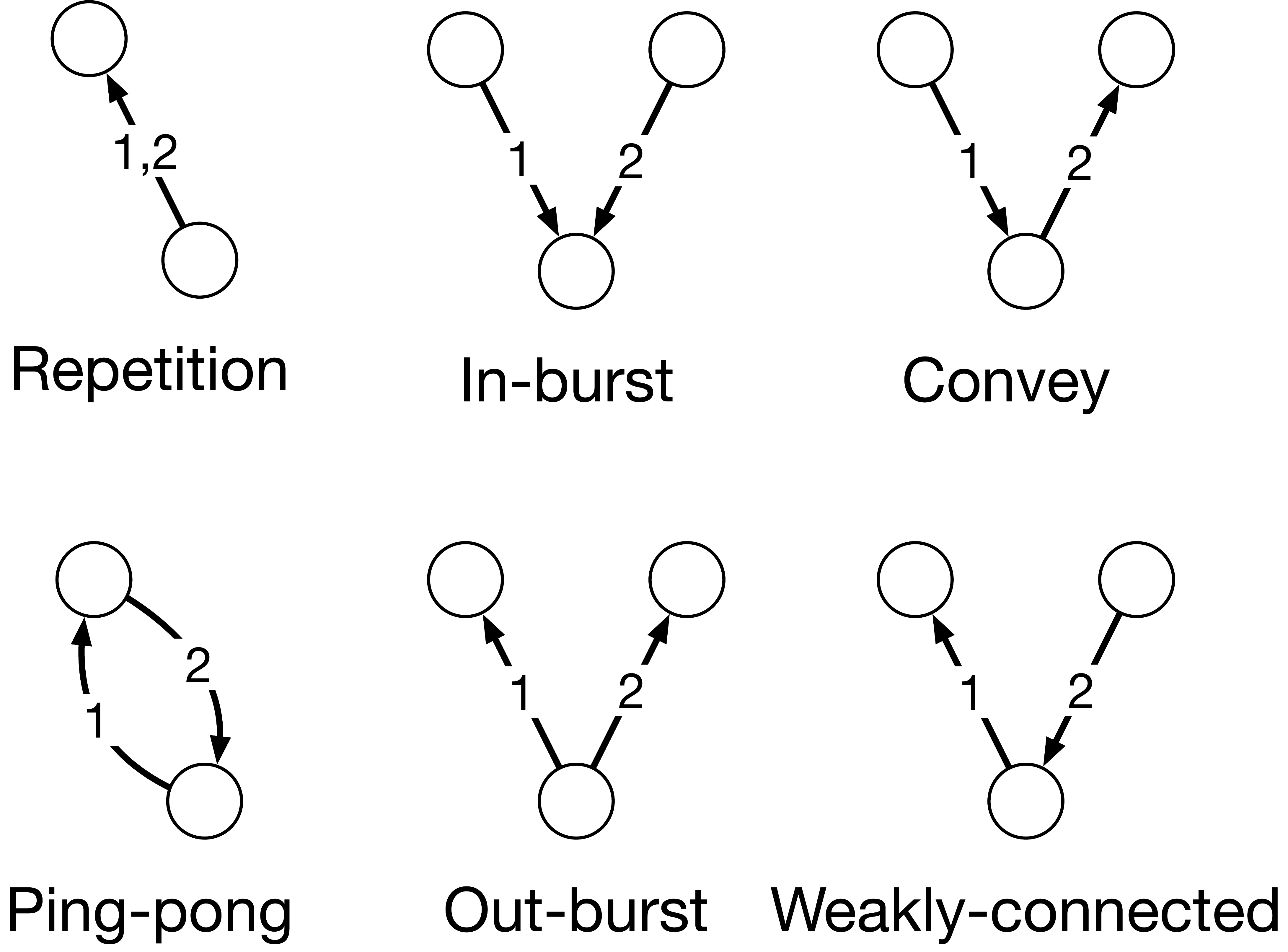} \\ 
(b)
\includegraphics[width=0.38\linewidth]{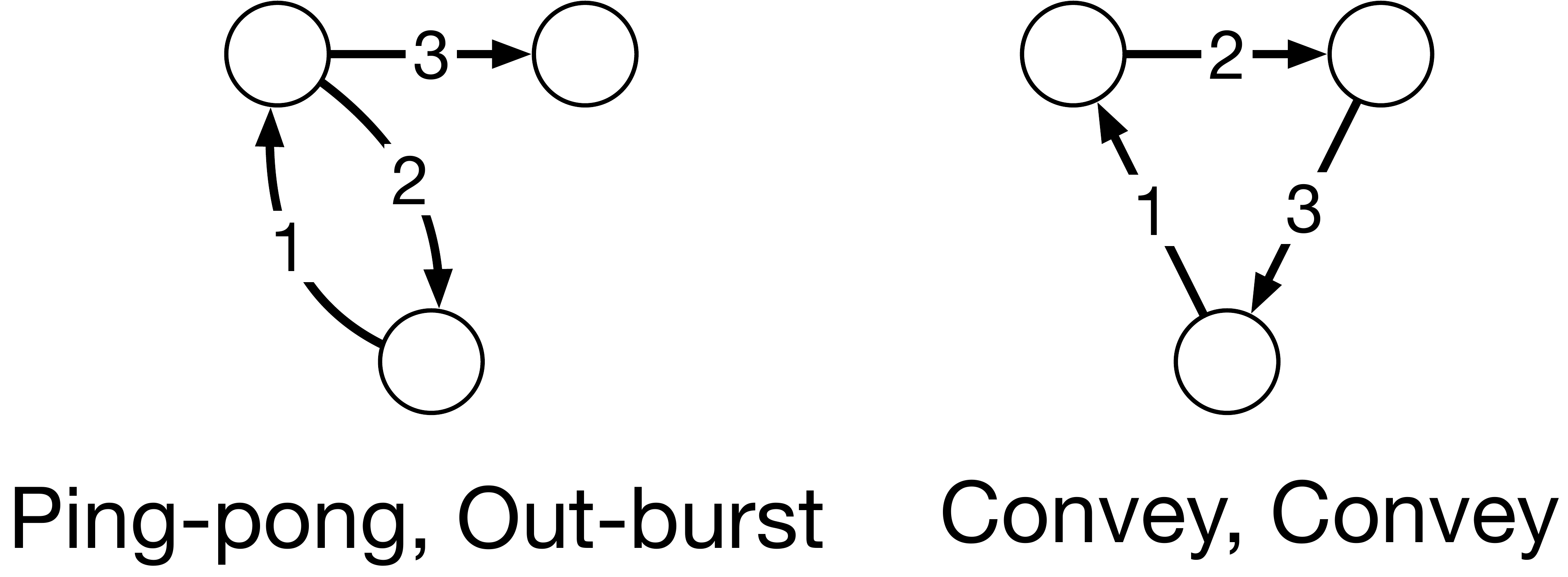} \\ 
\end{tabular}
&
(c)
\includegraphics[width=0.43\linewidth]{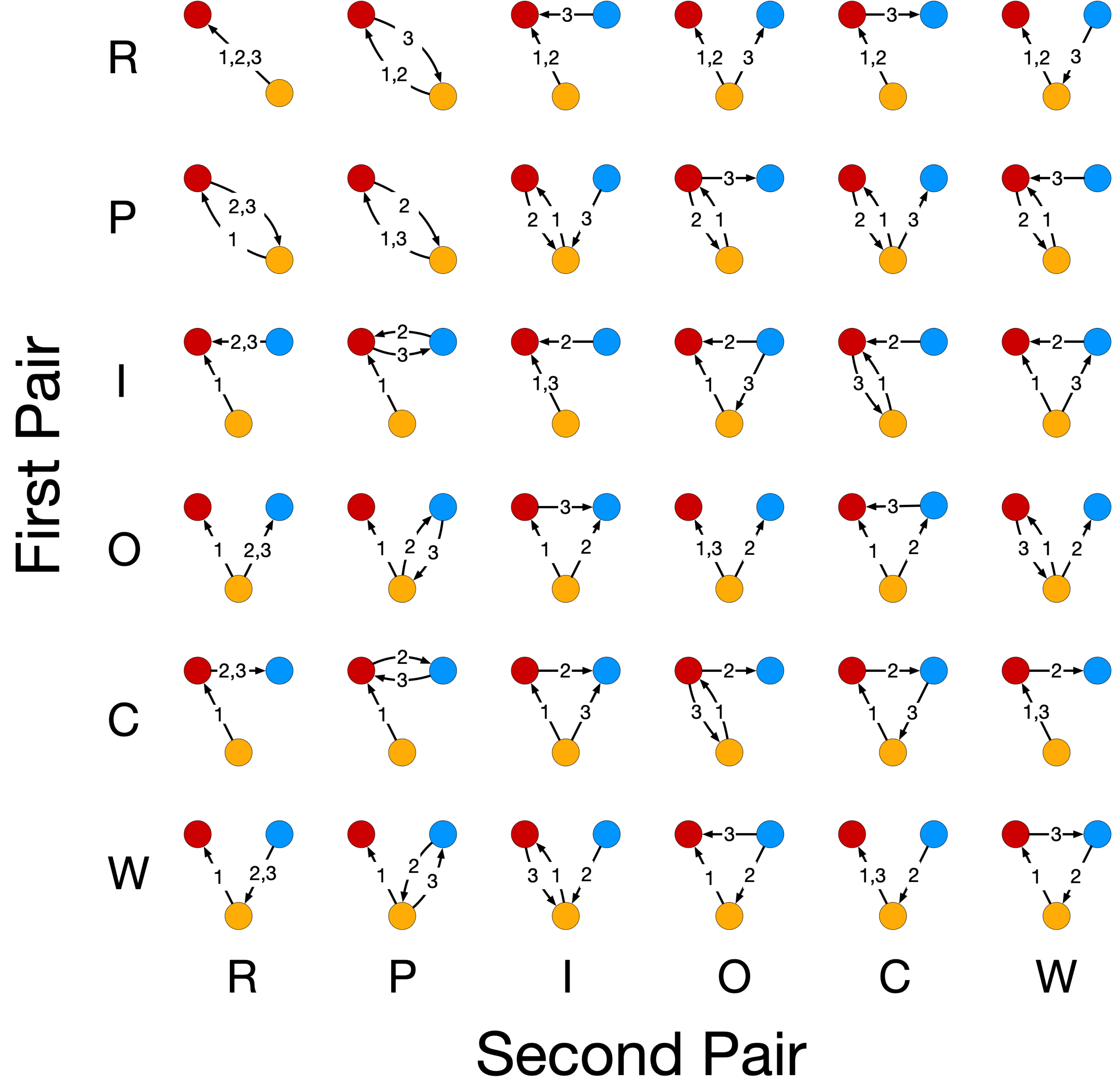} \\ 
\end{tabular}
\caption{\it (a) All 2-event motifs, which are time-ordered event pairs. (b) Two examples of 3-event motifs labeled by the sequence of two 2-event pairs. (c) All 2-node 3-event and 3-node 3-event motifs. R stands for repetition, P for ping-pong, I for in-burst, O for out-burst, C for convey, and W for weakly-connected.
}
\label{fig:notation}
\end{figure*}

A pair of events, i.e., 2-event motif, is the simplest temporal motif according to our definition.~\cref{fig:notation}(a) shows all types of 2-event motifs. Note that 2-node 2-event motifs are either a repetition, in which the two events occur in the same direction, or a ping-pong, in which the two events occur in the reciprocal directions. There are four types of 3-node 2-event motifs; in an in-burst motif, the two events share the same source; in an out-burst motif, the two events share the same target; in a convey motif, the source of the second event is the target of the first event; and in a weakly-connected motif, the target of the second event is the source of the first event.

A 3-event motif is composed of two event pairs, i.e., the pair composed of the first and second events and the pair composed of the second and third events. Therefore, each 3-event motif can be represented as a sequence of two event pairs. ~\cref{fig:notation}(b) gives two examples: a 3-node 3-event motif can be represented as a sequence of ping-pong and out-burst (denoted as P-O for convenience), while a 3-node cycle can be represented as a sequence of convey and convey (C-C). There are 4 types of 2-node 3-event motifs and 32 types of 3-node 3-event motifs (shown in ~\cref{fig:notation}(c)).
Among the 32 types of 3-node 3-event motifs, eight motifs are triangle motifs, of which the static projection when one ignores the direction of the edges is a closed triangle. The other 24 motifs are wedge motifs, of which the static projection is an open triangle (i.e., wedge). Among the triangle motifs, two of them are 3-node cycles, of which the static projection is a directed cycle, and the other six are acyclic triangle motifs.

\subsection*{Significance of temporal motifs} \label{sec2}
In static motif analyses, randomized networks are used as null models to measure the statistical significance of motifs. 
Milo et al.~\cite{milo2002network} analyzed the significance of motifs using the directed configuration model, which yields randomized networks that conserve the in-degree and out-degree of each node in the original network. One measures the statistical significance of motif $M_i$ by the $Z$ score defined as $Z_i = \frac{Original_i - \mu_i}{\sigma_i}$, where 
$Original_i$ is the number of $M_i$ motifs in the original data, and $\mu_i$ and $\sigma_i$ are the mean and standard deviation of the number of $M_i$ motifs computed from a given number of samples of randomized networks. 
For temporal networks, there are numerous reference models that randomize different structural or temporal aspects of the given temporal netowrk~\cite{gauvin2018randomized}.
We consider the following randomization methods to measure the $Z$ score for the temporal motifs: 

\begin{itemize}
\setlength\itemsep{0.1ex}
\setlength{\itemindent}{0.5ex}
\item \textit{Link shuffling (LS)} randomizes the edges using the Erd\H{o}s-R\'{e}nyi random graph model. The events and their timestamps are preserved for each edge.
\item \textit{Degree-constrained link shuffling (DCLS)} randomizes the edges using the directed configuration model, which conserves the in-degree and out-degree of each node. The events and their timestamps are preserved for each edge.
\item \textit{Weight-constrained timeline shuffling (WTS)} randomly generates the timestamps of all events for each edge. The number of events on each edge and the structure of the static network are preserved. The WTS corresponds to the assumption that a Poisson process generates the events on each edge, which is equivalent to randomly and independently assigning the timestamp of each event from the uniform distribution on the observation time window.
\item \textit{Inter-event shuffling (IS)} uniformly randomly shuffles the inter-event times on each edge while keeping the time of the first and the last event fixed. The number of events on each edge and the structure of the static network are preserved.
\item \textit{Timestamp shuffling (TS)} uniformly randomly shuffles the timestamps of all events regardless of the edges that the events belong to. The number of events on each edge and the structure of the static network are preserved.
\end{itemize}

LS and DCLS randomize edges and hence the structure of the static network, but not the temporal features. 
In contrast, WTS, IS, and TS randomize event times without changing the structure of the static network.

\section*{Results}
In this section, we examine temporal patterns in patent oppositions, collaborations, and the interplay between them. 
First, we find the temporal motifs in patent oppositions using different timing thresholds and investigate the sizes of the companies in the opposition motifs.
Second, we identify collaborations that exist between any pair of companies in the detected opposition motifs. Then, we analyze the interplay between patent oppositions and collaborations from both structural and temporal perspectives. 

\begin{figure*}[!b]
\centering
\includegraphics[width=1\linewidth]{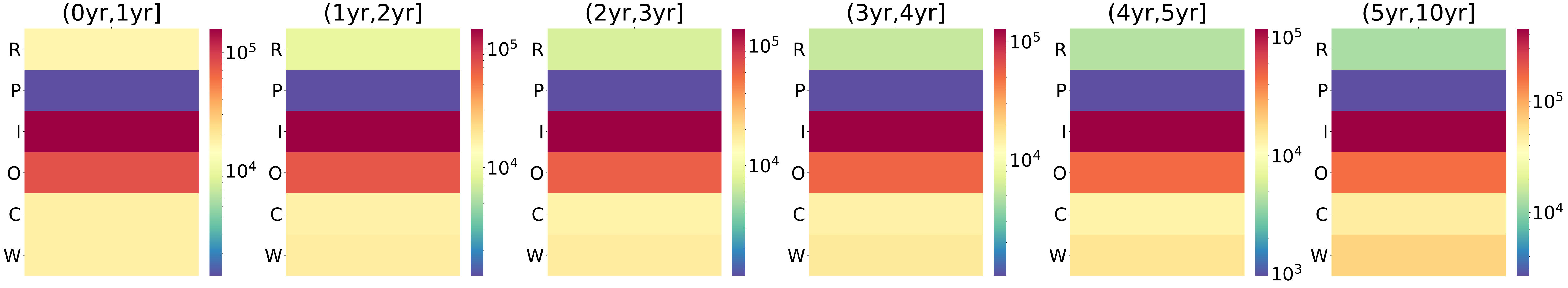}
\caption{\it The number of 2-event motifs in the opposition network using different timing thresholds.
Each row represents a type of 2-event motif; see the caption of~\cref{fig:notation} for the abbreviation. The color indicates the motif counts. For each heat map $(x, y]$, we only count the motifs of which the inter-event time is larger than $x$ and not greater than $y$.
}
\label{fig:real-2e}
\end{figure*}

\subsection*{Motifs in patent oppositions}
We first explore the temporal motifs in patent oppositions. In particular, we want to answer the following questions: 1) What are the frequent patterns in patent oppositions for a given timing threshold? 2) How does the size of the companies impact their roles in the patent oppositions?

\begin{figure*}[!h]
(a)
\includegraphics[width=\linewidth]{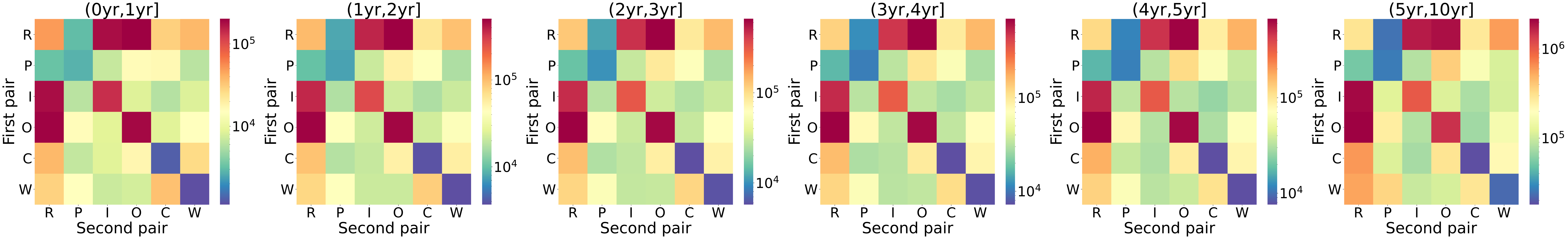}
(b)
\includegraphics[width=\linewidth]{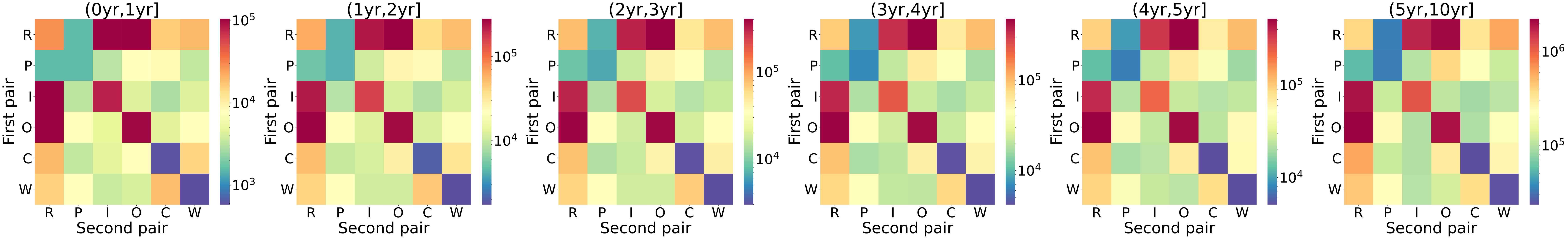}
\caption{\it The number of 2-node 3-event and 3-node 3-event motifs in the opposition network for a range of $\Delta_C$ values (a) or for a range of $\Delta_W$ values (b).
For each heatmap, $(x, y]$ indicates the values of the timing threshold, i.e., we only count the motifs for which $\Delta_C \in (x, y]$ and $\Delta_W=10$ years in panel (a), and $\Delta_C=\Delta_W \in (x, y]$ in panel (b).
Each cell in the heat map corresponds to a type of 3-event motif, where the row shows the first pair of events (i.e., the first two of the three events constituting the 3-event motif) and the column shows the second pair of events (i.e., the second and third events in the same motif). The color indicates the motif counts.
}
\label{fig:real}
\end{figure*}

\subsubsection*{Frequent patterns in patent oppositions}

We first investigate the number of 2-event motifs in the opposition network. Note that the two timing thresholds, $\Delta_C$ and $\Delta_W$, are equivalent for 2-event motifs since the inter-event time of a 2-event motif is equal to its timespan. We select six timing thresholds and count the numbers of motifs for each timing threshold.
The results shown in~\cref{fig:real-2e} indicate that the repetition motifs are more abundant than the ping-pong motifs.
This result indicates that patent oppositions are more like one-way attack rather than a fight between two; a company may consistently oppose its opponent many times, but a company being opposed relatively infrequently opposes back.
In-burst and out-burst motifs are the most frequently observed among the six types of 2-event motifs. This result suggests that companies tend to oppose or be opposed by multiple companies within a short time.
Regarding the motif counts across different timing thresholds, in-burst and out-burst motifs are consistently the most abundant.
Then comes the convey motifs, where an opposed company copies the behavior to hinder others, and weakly-connected motifs, where an opposer company triggers an opposition to itself.
Repetition and ping-pong motifs are the least abundant but their fractions among the 2-event motifs are slightly larger for small timing thresholds.
The ranking of the six types of motif counts is the same for all the timing thresholds, except $(0yr, 1yr]$ where the weakly-connected motifs are observed less frequently than the convey motifs.

\begin{figure*}[!b]
\centering
(a)
\includegraphics[width=0.43\linewidth]{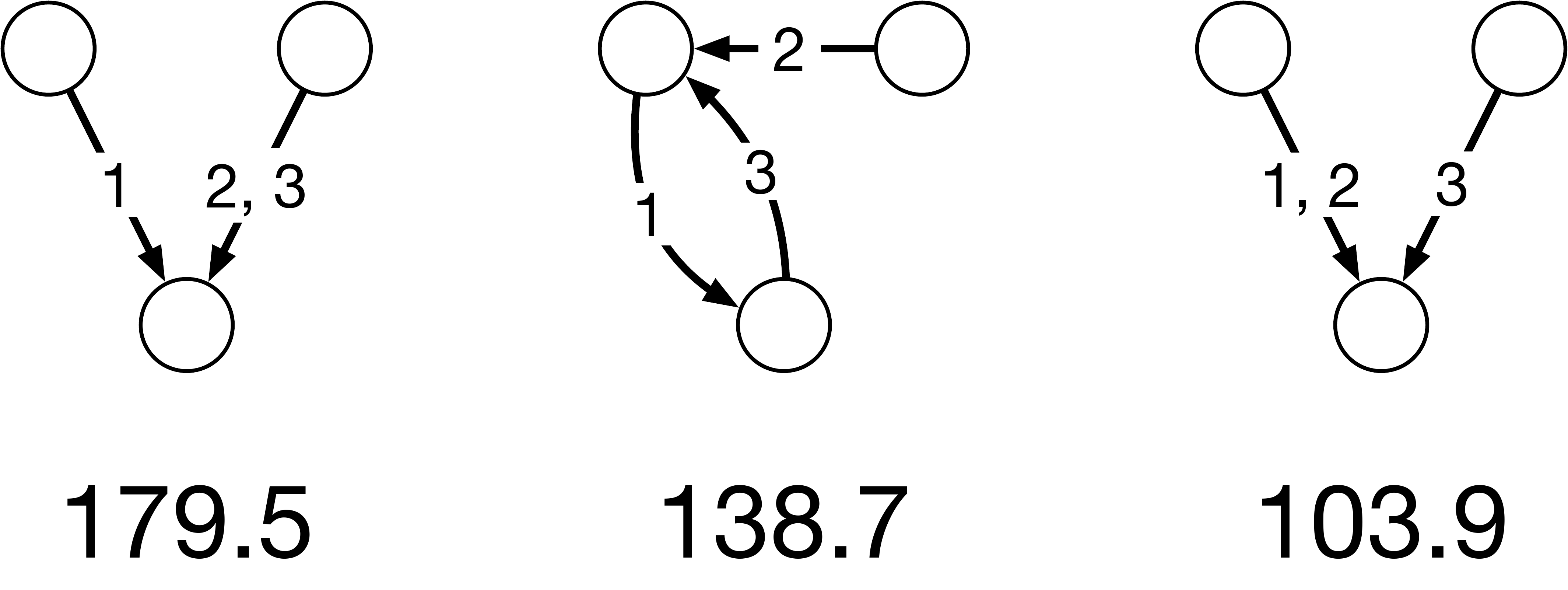}
\hspace{2ex}
(b)
\includegraphics[width=0.43\linewidth]{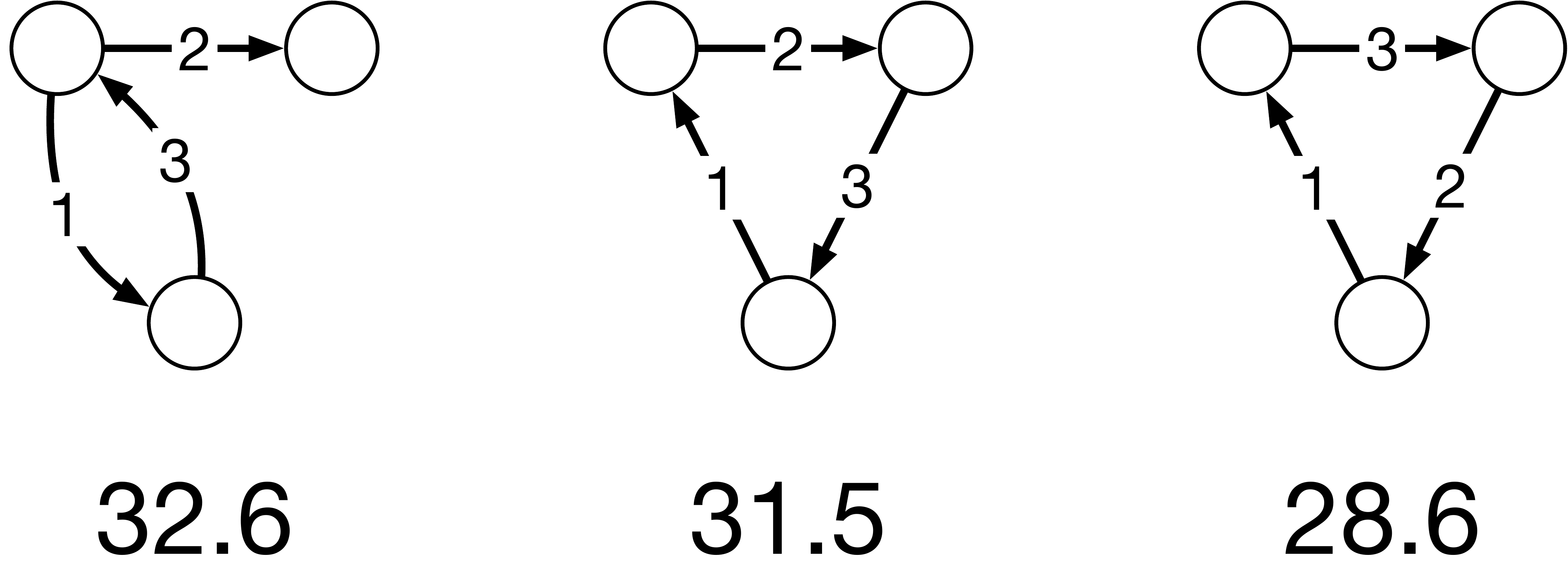}
\caption{\it The three most significant (a) and the three least significant (b) 3-event motifs based on the $Z$ score. The numbers shown are the Z scores. We used the WTS model as null model and set $\Delta_W = 10$ years.}
\label{fig:z-score}
\end{figure*}

Next, we count the numbers of 2-node 3-event and 3-node 3-event motifs for $\Delta_C$, $\Delta_W$$= 1, 2, 3, 4, 5,$ and $10$ years. 
We first set $\Delta_W=10$ years and consider a range of $\Delta_C$ values.
Then, we set $\Delta_C=\Delta_W$ for a range of $\Delta_W$ values (i.e., we only consider the time-window threshold and do not enforce the inter-event timing threshold).
The numbers of motifs of each case are shown in \cref{fig:real}. We observe the following patterns consistently for the different timing threshold values. First, the motifs containing in-bursts and out-bursts are dominant. Second, motifs containing repetitions are more abundant than those containing ping-pongs.
Third, the 2-node 3-event motifs (i.e., R-R, R-P, P-R, P-P) are progressively less abundant than the 3-node 3-event motifs when $\Delta_C$ or $\Delta_W$ is larger.
These three results are consistent with the results for 2-event motifs shown in~\cref{fig:real-2e}.
Fourth, triangle motifs are relatively rare. There are eight types of triangle motifs but they only occupy 3\% of the total motif count when $\Delta_C = $ 1 year, for example. The last result is coherent with the structural balance theory~\cite{heider1946attitudes, cartwright1956structural}. Because patent oppositions are negative relations, the triangle of oppositions is an imbalanced structure.
In addition, the 3-node cycles, i.e., C-C and W-W motifs, appear less frequently than the other six acyclic triangle motifs. A similar pattern has been found in static social networks where a negative edge is less likely to occur with two other negative edges in a cycle~\cite{leskovec2010signed}.

In addition to the motif counts, we measure the statistical significance of each type of opposition motif using different null models.
To this end, we generate 10 random graphs for each null model and compute the $Z$ score for each type of opposition motif (see Supplementary Table 1).
For the LS and DCLS null models, some motifs are not observed at all or have zero variation across different realizations of the randomized network.
The Z scores are positively large, safely significantly so after controlling for multiple comparison (i.e., > 10), for all the 2-node 2-event motifs and 3-node 2-event motifs for the WTS and TS.
\cref{fig:z-score} shows the most and the least significant opposition motifs using the WTS as null model and $\Delta_W= 10$ years.
The WTS model uniformly randomly generates the timestamps for all events assuming that events are generated by a Poisson process on each edge.
The three most significant motifs include wedge motifs that contain in-bursts and repetitions (I-R, W-I, R-I), while the three least significant motifs contain two 3-node cycles (O-W, C-C, W-W). 
These results justify our analysis stated earlier in this section based on the motif counts: the motifs containing repetitions and in-bursts tend to be frequent, whereas the cycle motifs are relatively scarce.

\begin{figure*}[t!]
\centering
\includegraphics[width=0.8\linewidth]{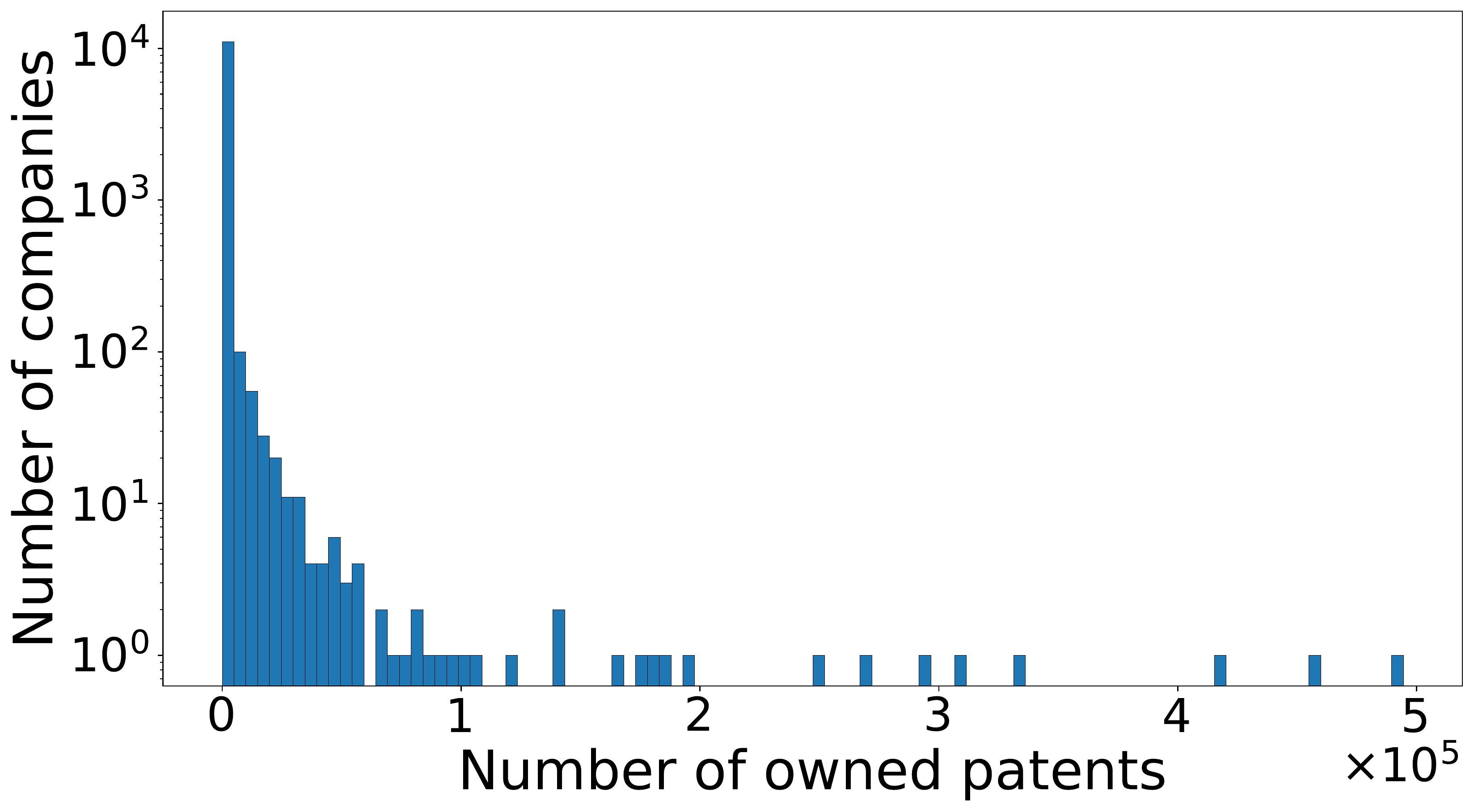}
\caption{\it The distribution of the number of patents owned by the companies.
}
\label{fig:size}
\end{figure*}

\begin{table*}[!b]
\centering
\begin{tabular}{|m{1.2 cm}||M{1.2 cm}|M{1.2 cm}|M{1.2 cm}||M{1.2 cm}|M{1.2 cm}|M{1.2 cm}||M{1.2 cm}|M{1.2 cm}|M{1.2 cm}|}
\hline
\multicolumn{1}{|c||}{\multirow{2}{*}{Motif}}    & \multicolumn{3}{c||}{\includegraphics[width=0.09\linewidth]{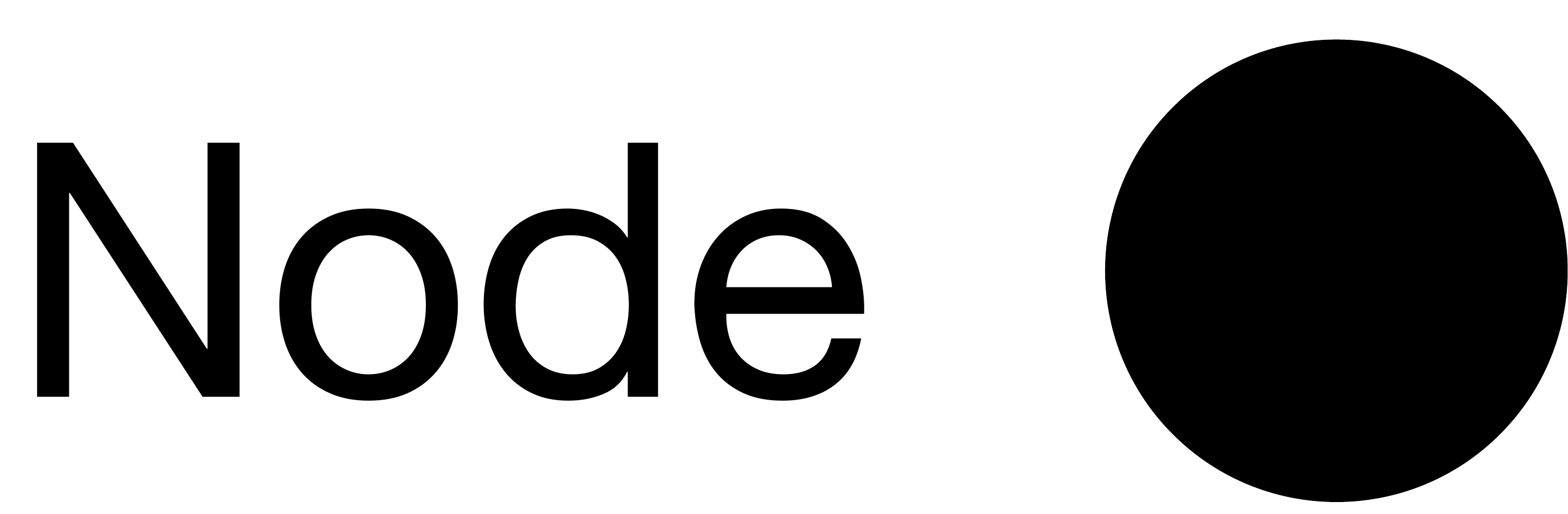}}                 & \multicolumn{3}{c||}{\includegraphics[width=0.09\linewidth]{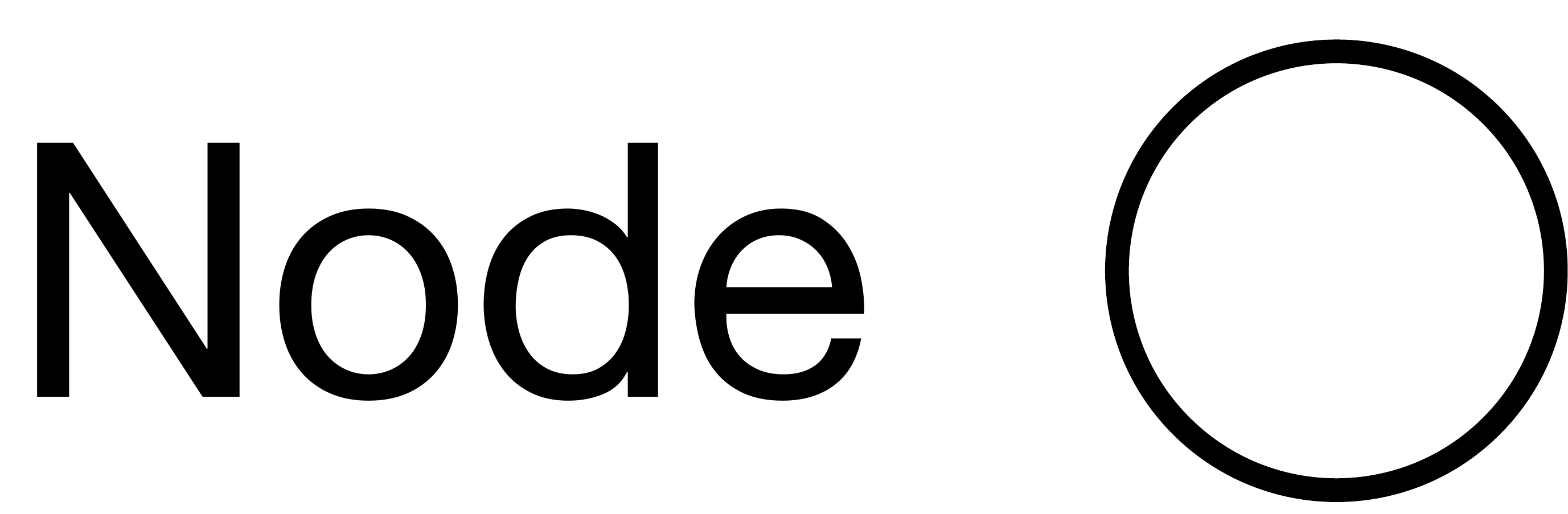}}                 & \multicolumn{3}{c|}{\includegraphics[width=0.09\linewidth]{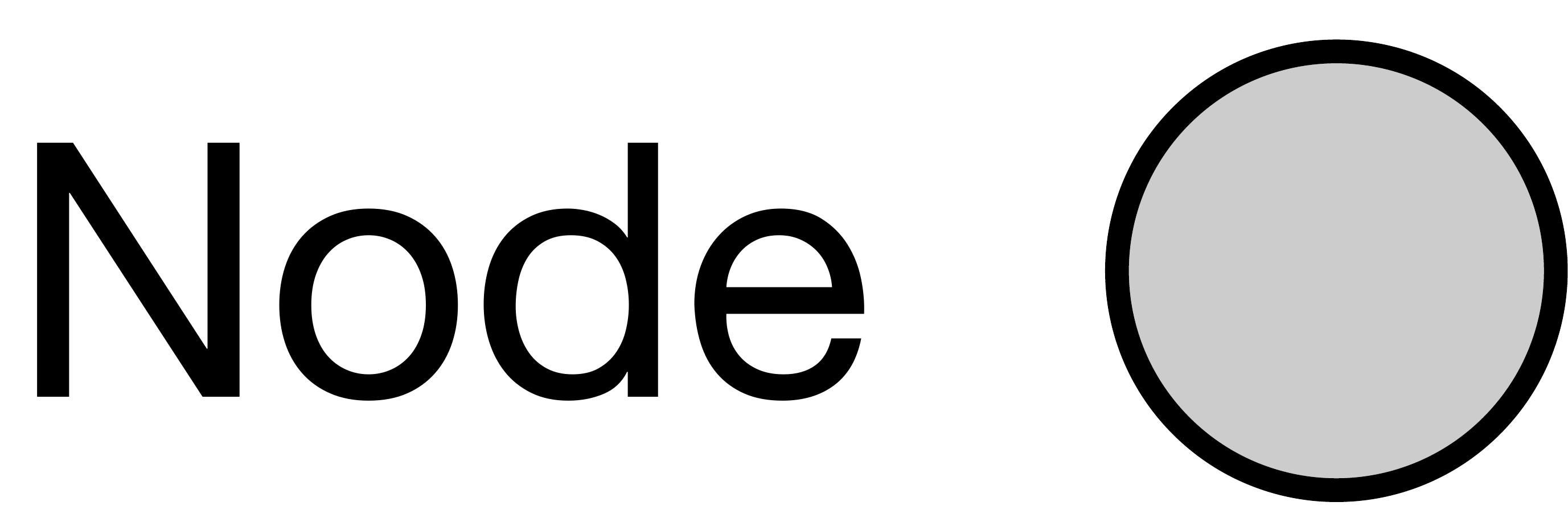}}                 \\ 
 & Mean & Median & Std & Mean  & Median & Std & Mean  & Median & Std \\  \hline
\vspace{0.1ex}\includegraphics[width=1 cm]{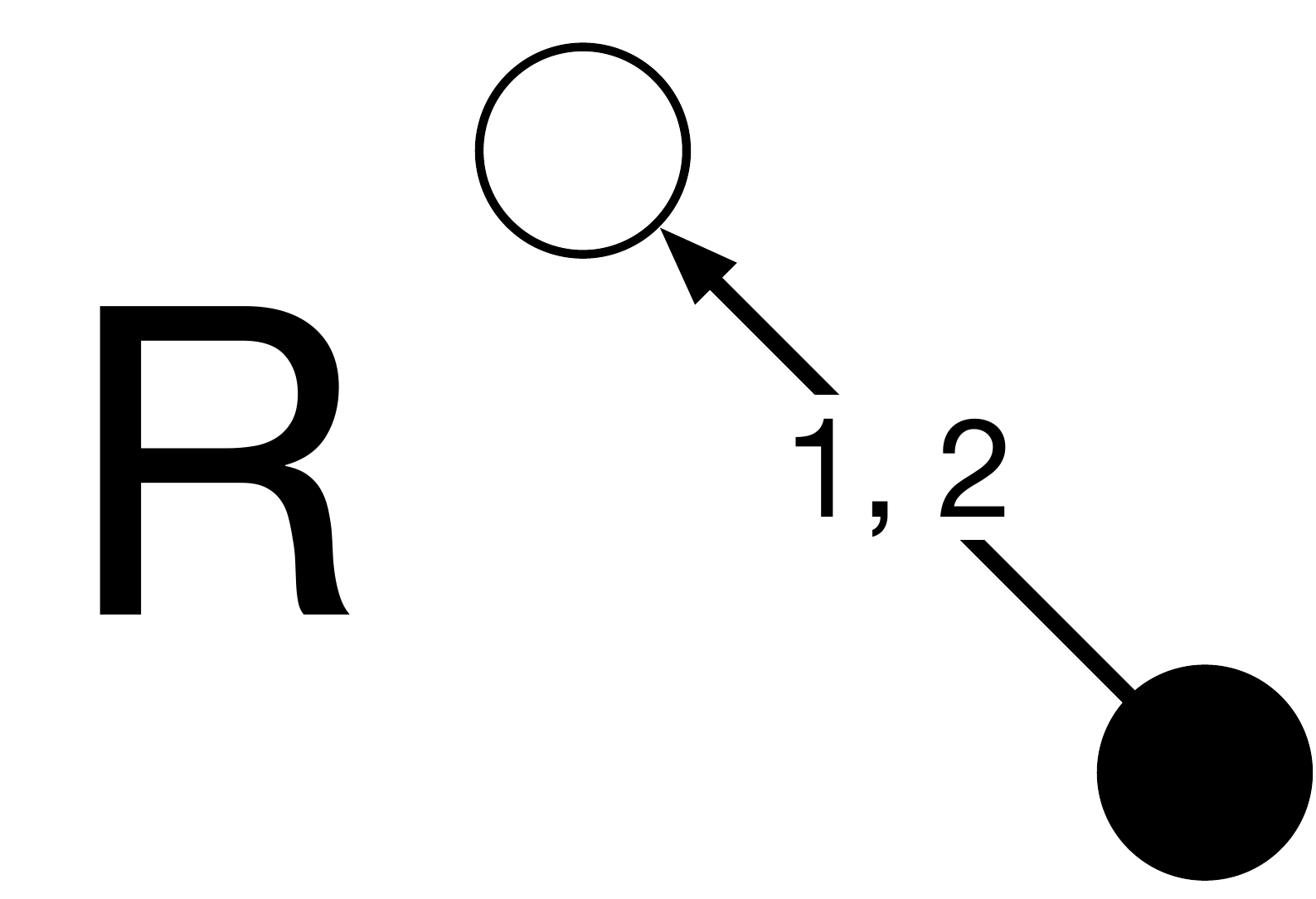}\vspace{-0.3ex}	&	9227.7	&	6523	&	19142.0	&	8943.5	&	2232	&	23997.9	&	\multicolumn{3}{c|}{N/A} \\ \hline			
\vspace{0.1ex}\includegraphics[width=1 cm]{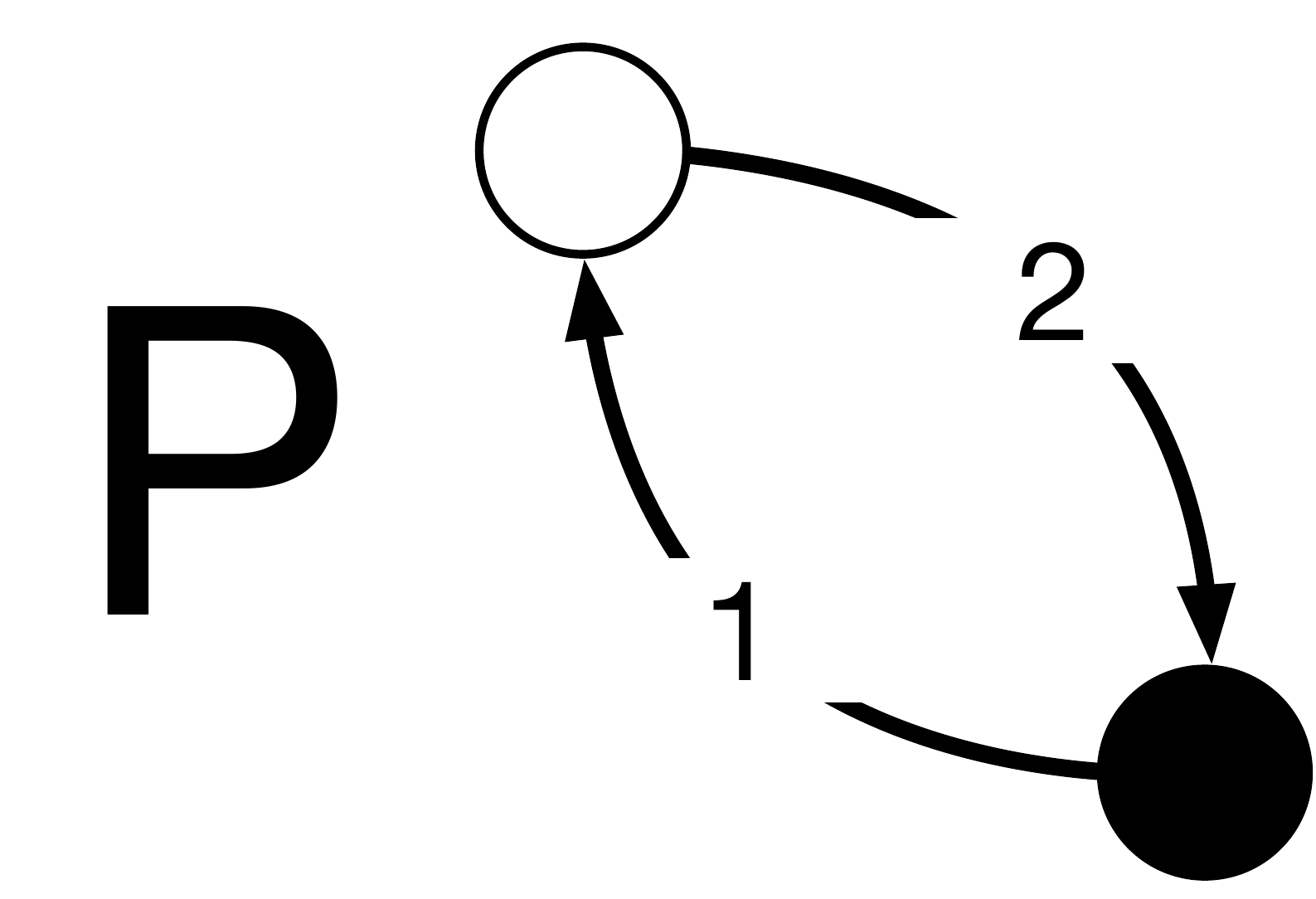}\vspace{-0.3ex}	&	7305.7	&	5812	&	8598.0	&	7703.5	&	6523	&	10469.7	&	\multicolumn{3}{c|}{N/A} \\ \hline				
\vspace{0.1ex}\includegraphics[width=1 cm]{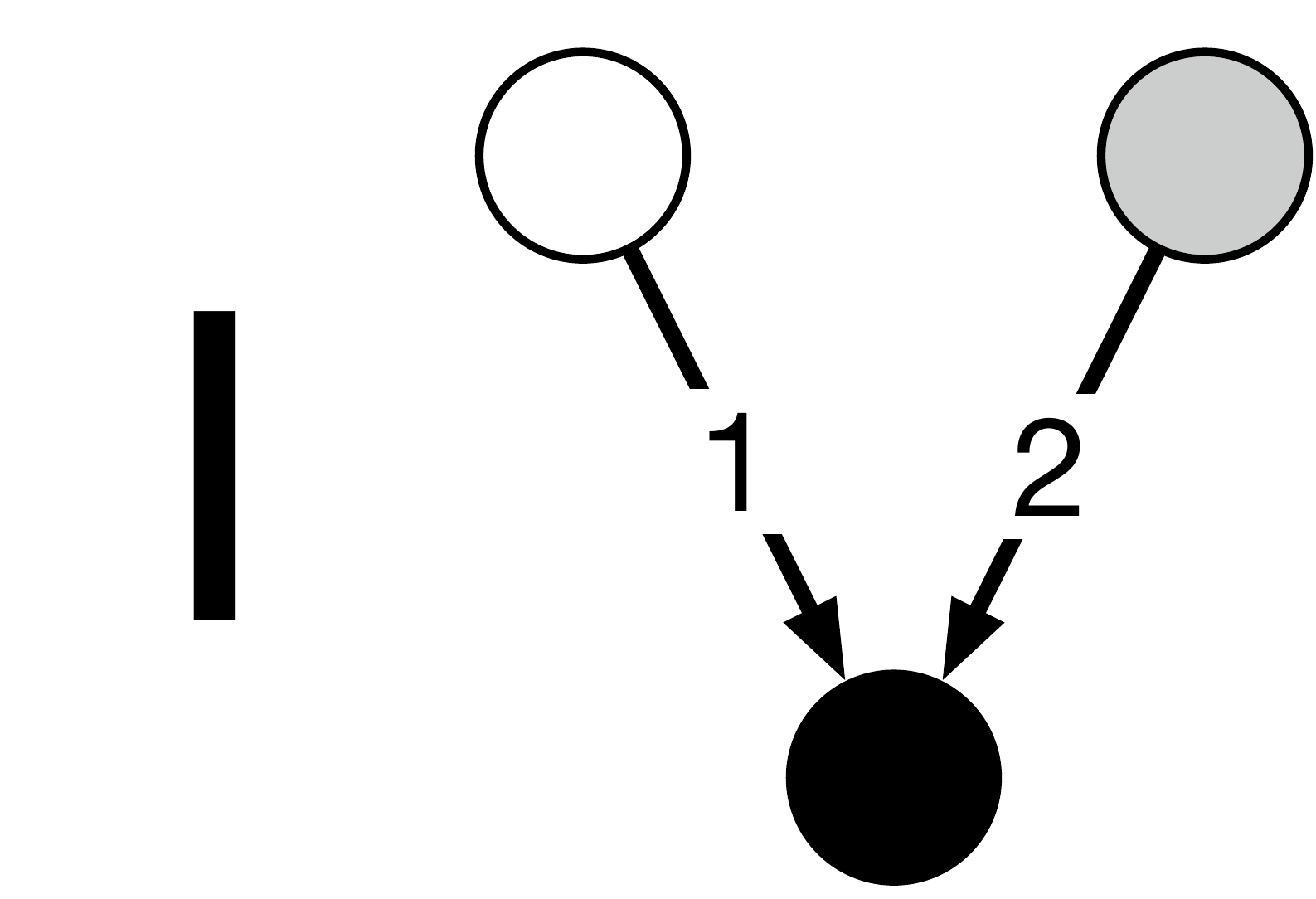}\vspace{-0.3ex}	&	5194.5	&	342	&	24176.1	&	8102.8	&	394	&	28014.7	&	4867.6	&	248	&	18135.4	 \\ \hline
\vspace{0.1ex}\includegraphics[width=1 cm]{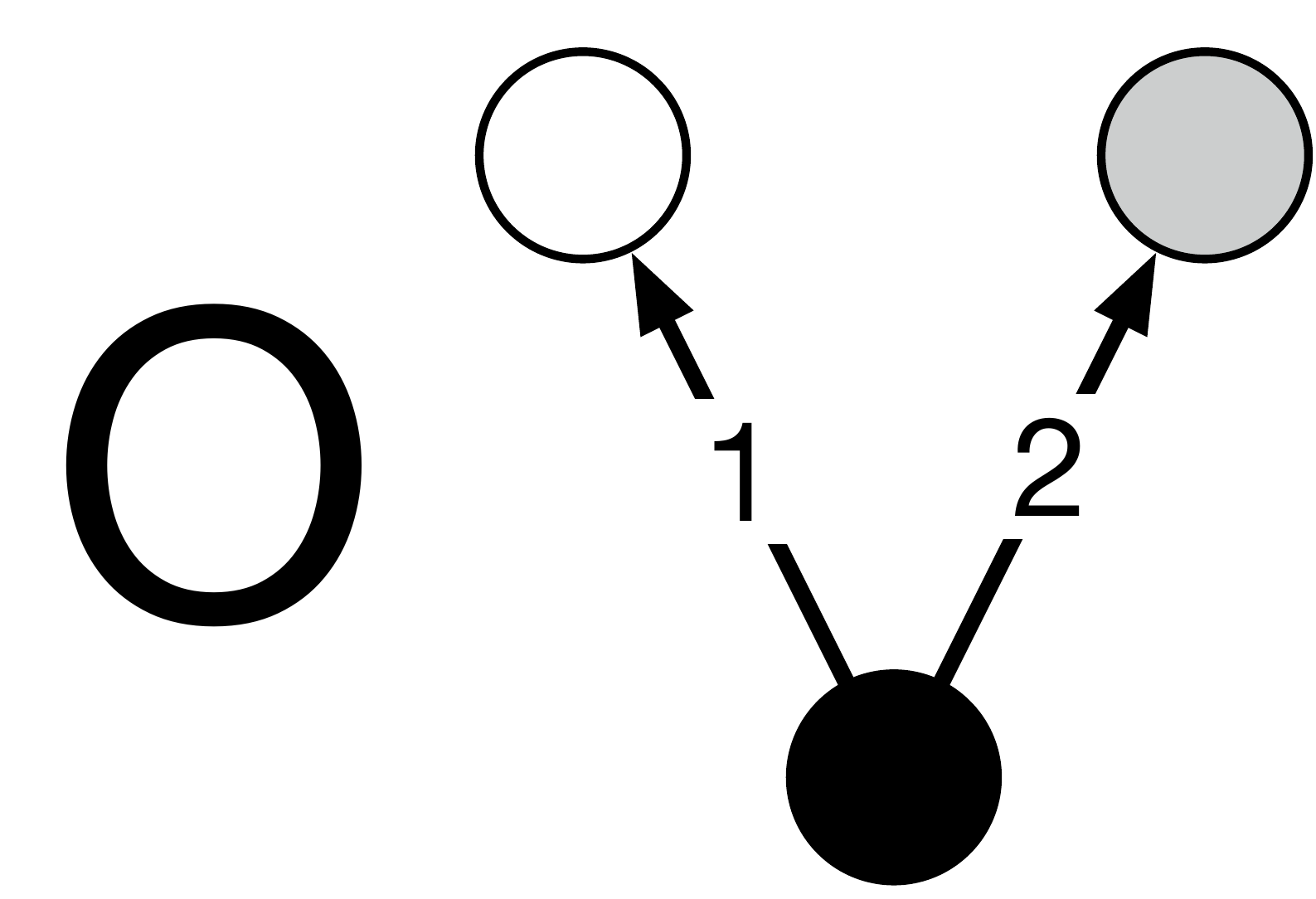}\vspace{-0.3ex}	&	18388.7	&	12313	&	39151.5	&	8625.8	&	493	&	35229.4	&	9638.1	&	775	&	36575.8	 \\ \hline
\vspace{0.1ex}\includegraphics[width=1 cm]{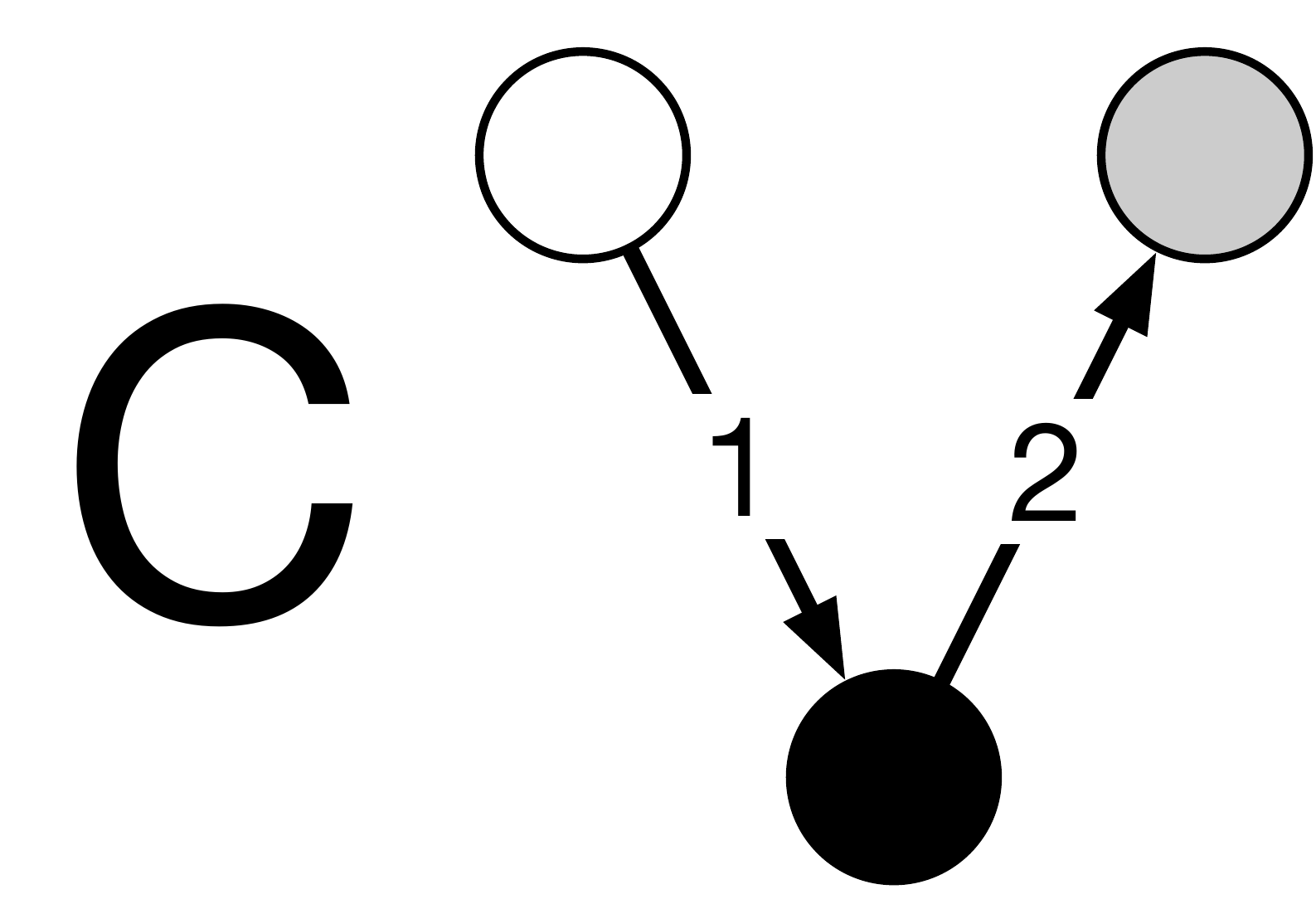}\vspace{-0.3ex}	&	16064.3	&	12313	&	30410.0	&	3949.6	&	479	&	9403.3	&	9111.7	&	1195	&	31176.3	 \\ \hline
\vspace{0.1ex}\includegraphics[width=1 cm]{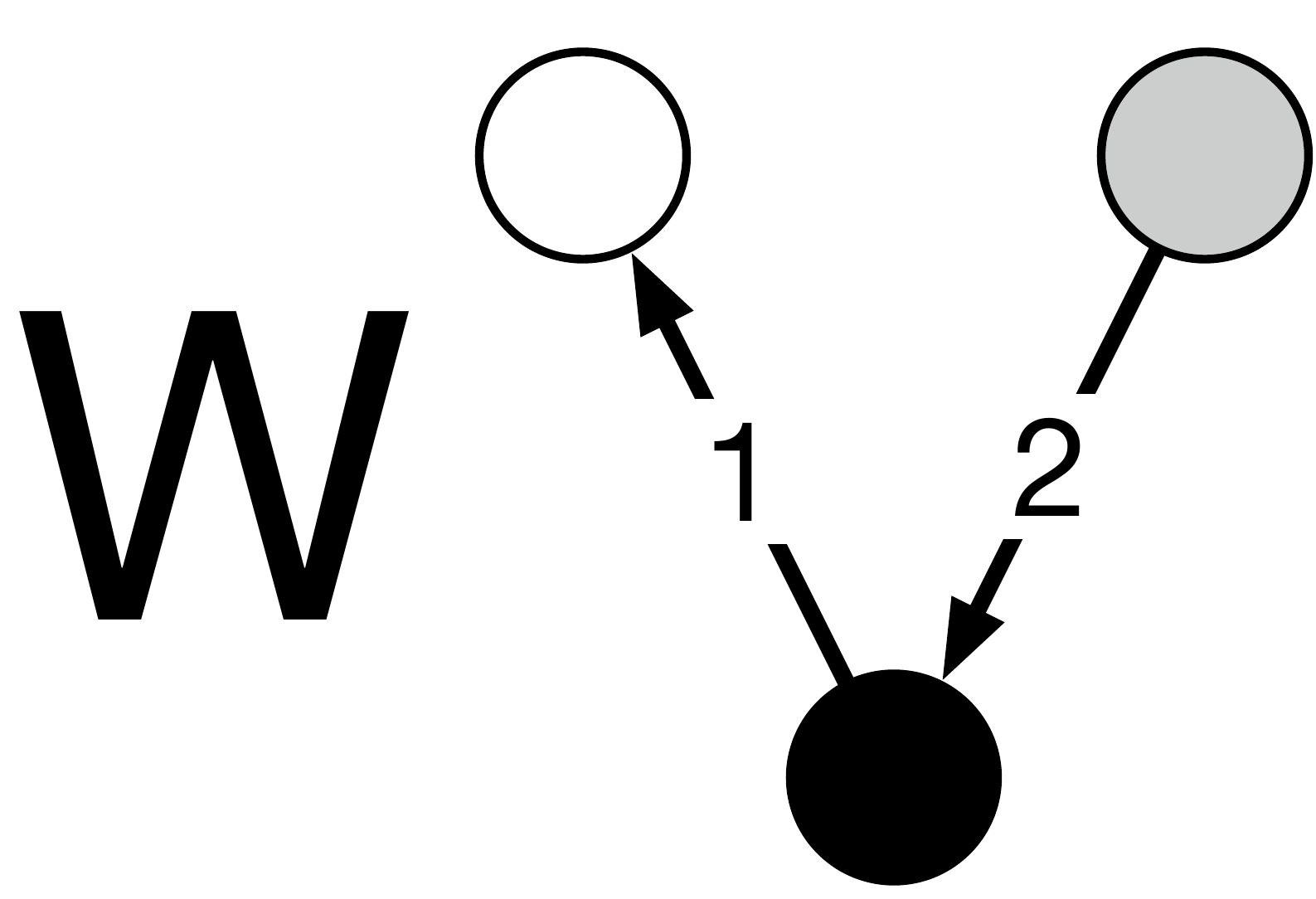}\vspace{-0.3ex}	&	20309.1	&	12313	&	39205.7	&	8638.2	&	503	&	35458.6	&	3606.2	&	354	&	9143.4	 \\ \hline
\end{tabular}
\caption{\it Size of the company for each node in the 2-event opposition motifs. We show the mean, median, and standard deviation (denoted by Std in the table) of the number of patents owned by the companies for each node in a given type of motif. 
For the entire network, the opposer and opposed companies have 4119.9 and 6492.8 patents on average, respectively.}
\label{tab:size-temporal}
\end{table*}

\subsubsection*{Position of companies of different sizes in the opposition motifs}

Companies of different sizes may participate in temporal network motifs by occupying particular positions in the motif. To explore this possibility, we examine the relationship between the sizes of the companies and their positions within the opposition motifs. 
For simplicity, we define the size of a company by the number of patents that the company owns. Note that this is often much greater than the number of patents involved in the oppositions.
The mean and median of the numbers of patents owned by the companies are 959.07 and 25, respectively, while the minimum and maximum are 0 and 494,463, respectively.
We show the distribution of the number of owned patents in~\cref{fig:size}.
We identify all 2-event motifs with $\Delta_C = \Delta_W = $ 10 years in the opposition network. 

For each position in each type of opposition motif, we show the mean, median, and the standard deviation of the number of patents owned by those companies in~\cref{tab:size-temporal}.
Considering the opposer and opposed company pairs in all the opposition events, the mean of the number of patents owned by the opposers is 4119, the median is 162, and the standard deviation is 18330. For opposed companies, the mean of the number of patents owned is 6492, the median is 214, and the standard deviation is 31427.
Although the mean size of the company does not significantly depend on the position in the motif due to a relatively large standard deviation, we observe the following tendencies.
First, the result for repetition and ping-pong motifs suggests that the two companies forming the motif tend to have similar sizes.
In the in-burst motifs, the opposed company is smaller than the first opposer but larger than the second opposer.
In the out-burst motifs, the opposer company is larger than the opposed companies.
In the convey and weakly-connected motifs, the center company (which opposes and is opposed) is substantially larger than the other two companies.
In almost all positions, the companies that are involved in opposition motifs are larger than the average company in the network.
Second, in most motifs, the opposer companies are larger than the average opposer in the network, with the exception of first opposers in convey and weakly-connected motifs.
The opposer company in the out-burst motifs and the center company in the convey and weakly-connected motifs tend to be particularly large.
Third, among the companies which only oppose one company and are not opposed by any company within the motif (i.e., the opposers in the in-burst and each opposer in the convey and weakly-connected motifs), the opposers in the in-burst motifs tend to be larger than the other two.
Interestingly, the first opposer in the in-burst motif tends to be larger than the second opposer.
The companies which are only opposed by one company and do not oppose any company within the motif (i.e., the two opposed companies in the out-burst and each opposed company in the convey and weakly-connected motifs) have similar sizes.
Fourth, among the companies that oppose twice and are not opposed by any, the opposer company in the out-burst motifs is substantially larger than the opposer in the repetition motifs on average.
Among the companies that are opposed twice and do not oppose any, the opposed company in the in-burst motifs is smaller than the opposed company in the repetition motifs.
These results suggest that the large companies tend to oppose multiple companies rather than the same company multiple times and that they are more likely to be opposed by one company multiple times than opposed by multiple companies.
Lastly, among the companies that oppose only one company and is opposed by only one company (i.e., both companies in the ping-pong and the center companies in the convey and weakly-connected motifs), the ones in the convey and weakly-connected are substantially larger than the companies in the ping-pong motifs on average.
This result suggests that the large companies tend to engage with multiple companies rather than the same company multiple times.

To validate whether our findings are unique to temporal motifs, we also carry out a similar analysis for static motifs in the static network of patent oppositions. We show the results in~\cref{tab:size-static}. 
For each position in each motif, the standard deviation of the company size in static motifs is considerably larger than in temporal motifs. This result suggests that the tendencies in the static motifs are less consistent. 
Another observation is that the size of the companies in static in-burst motifs shows a different behavior than in temporal motifs: the opposed company tends to be larger than the other two opposer companies.
In sum, the tendencies of company sizes we observed in the temporal motifs are not just consequences of the structure of the static networks, but rather are affected by the temporal relationships among patent oppositions.

\begin{table*}[!t]
\centering
\begin{tabular}{|m{1.2 cm}||M{1.2 cm}|M{1.2 cm}|M{1.2 cm}||M{1.2 cm}|M{1.2 cm}|M{1.2 cm}||M{1.2 cm}|M{1.2 cm}|M{1.2 cm}|}
\hline
\multicolumn{1}{|c||}{\multirow{2}{*}{Motif}} & \multicolumn{3}{c||}{\includegraphics[width=0.09\linewidth]{figs-final/node-black.pdf}}                 & \multicolumn{3}{c||}{\includegraphics[width=0.09\linewidth]{figs-final/node-white.pdf}}                 & \multicolumn{3}{c|}{\includegraphics[width=0.09\linewidth]{figs-final/node-grey.pdf}}                 \\ 
 & Mean & Median & Std & Mean  & Median & Std & Mean  & Median & Std \\  \hline
\vspace{0.3ex}\includegraphics[width=1 cm]{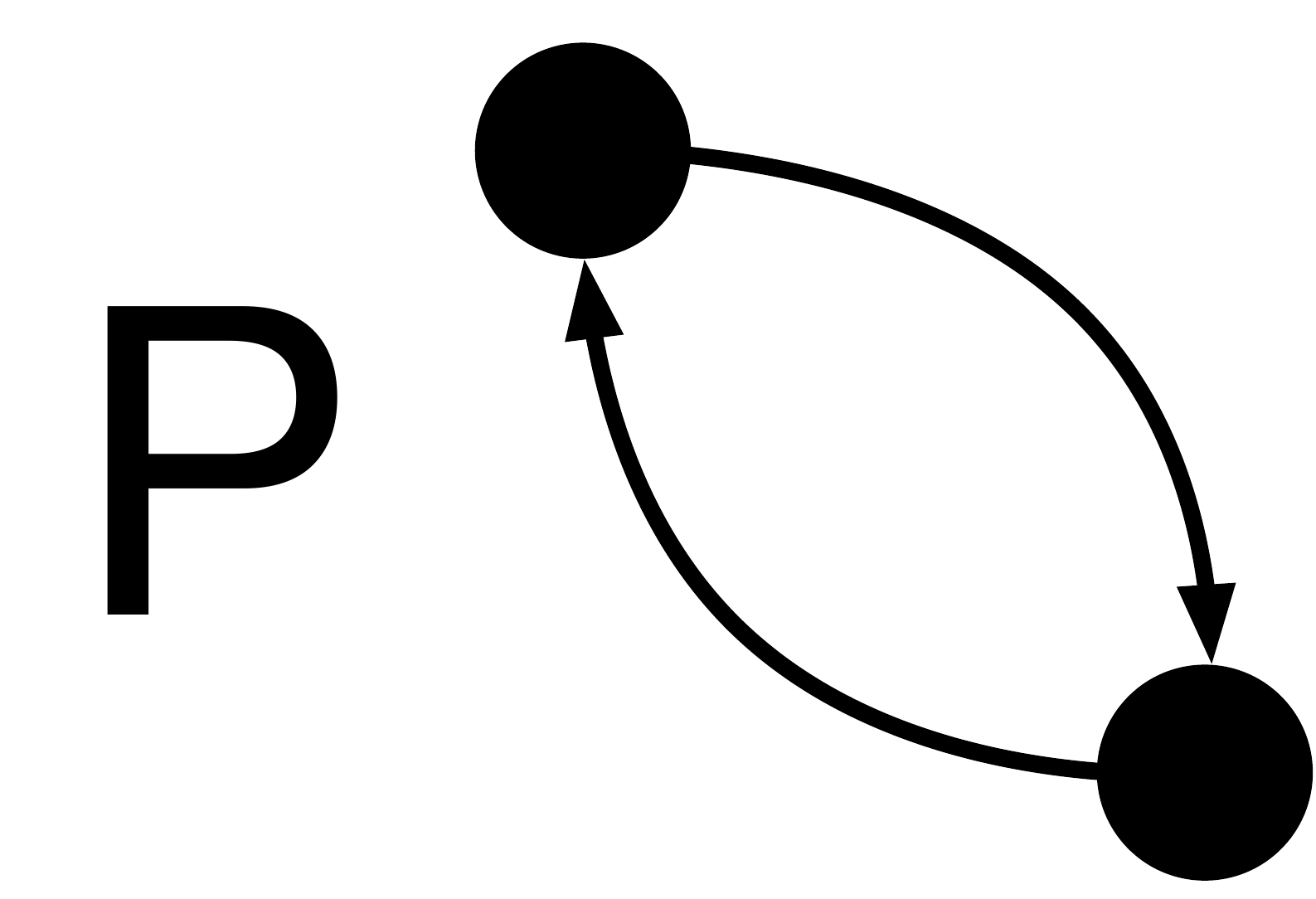}\vspace{-0.3ex}	&	4127.6	&	431.5	&	15589.5	&	\multicolumn{3}{c||}{N/A}	&	\multicolumn{3}{c|}{N/A}  \\ \hline			
\vspace{0.3ex}\includegraphics[width=1 cm]{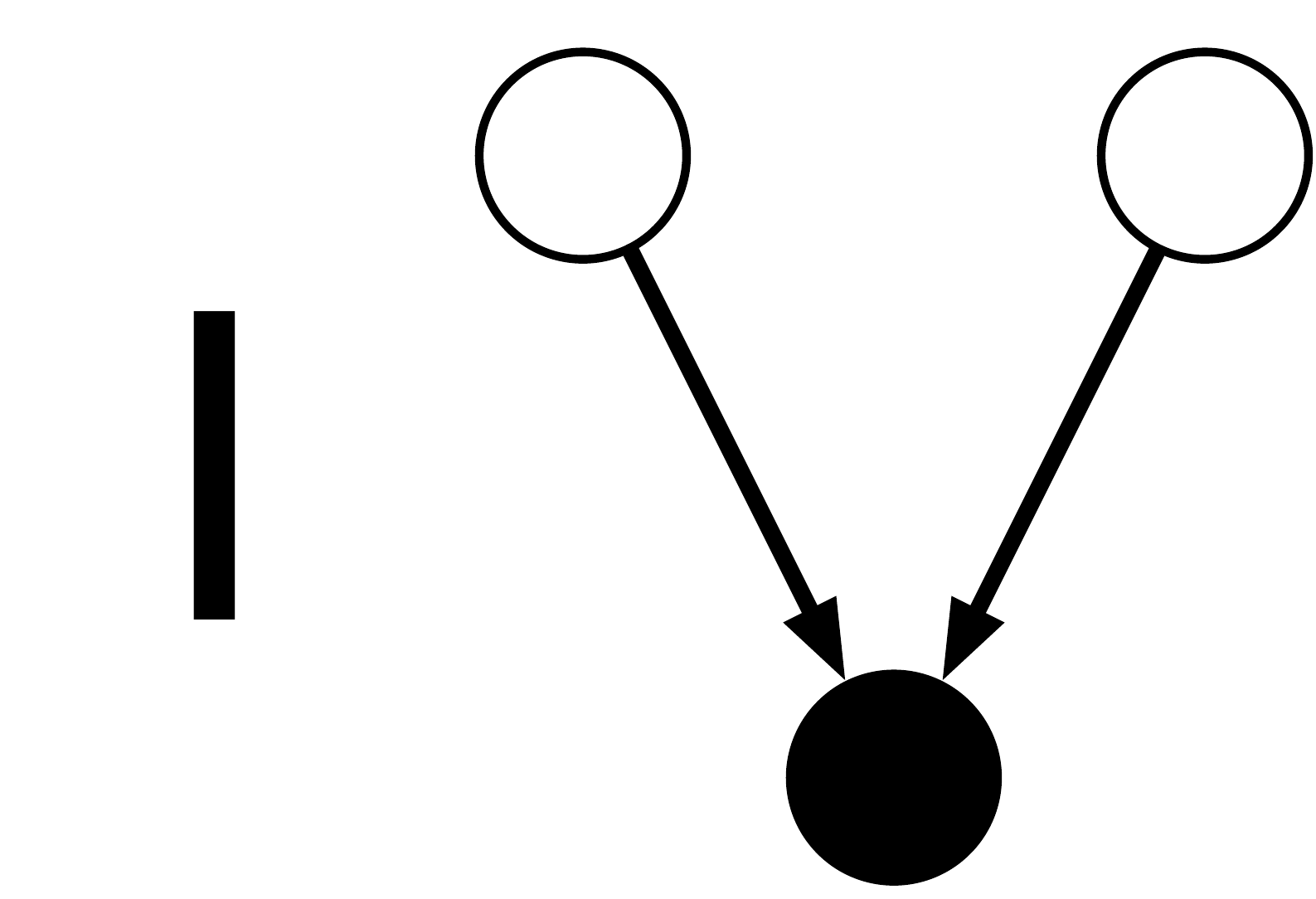}\vspace{-0.3ex}	&	7939.4	&	329	&	33576.4	&	2924.9	&	128	&	14603.6	&	\multicolumn{3}{c|}{N/A} \\ \hline		
\vspace{0.3ex}\includegraphics[width=1 cm]{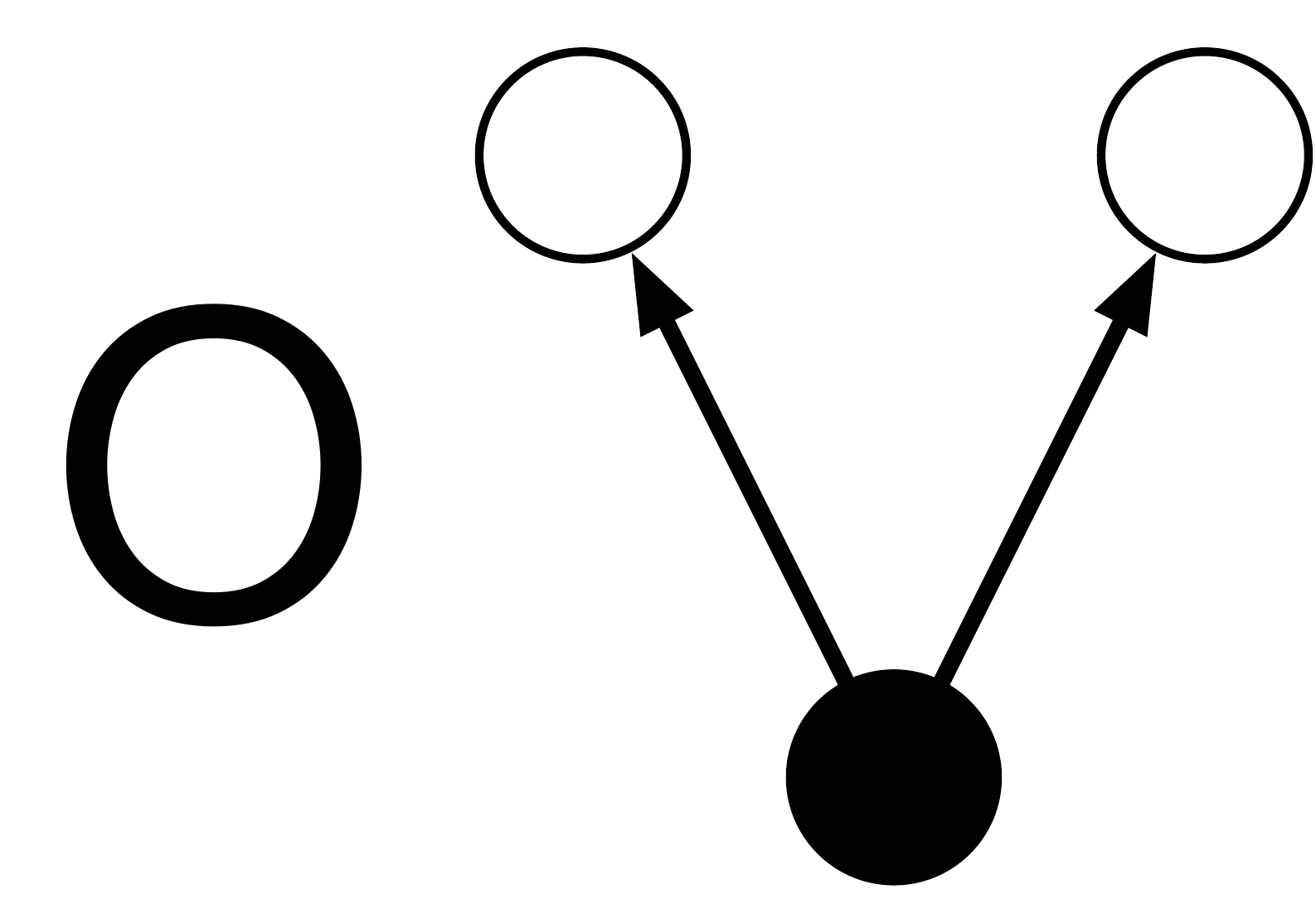}\vspace{-0.3ex}	&	18812.1	&	2492	&	45090.4	&	8490.0	&	202	&	41724.4	&	\multicolumn{3}{c|}{N/A}  \\ \hline	
\vspace{0.3ex}\includegraphics[width=1 cm]{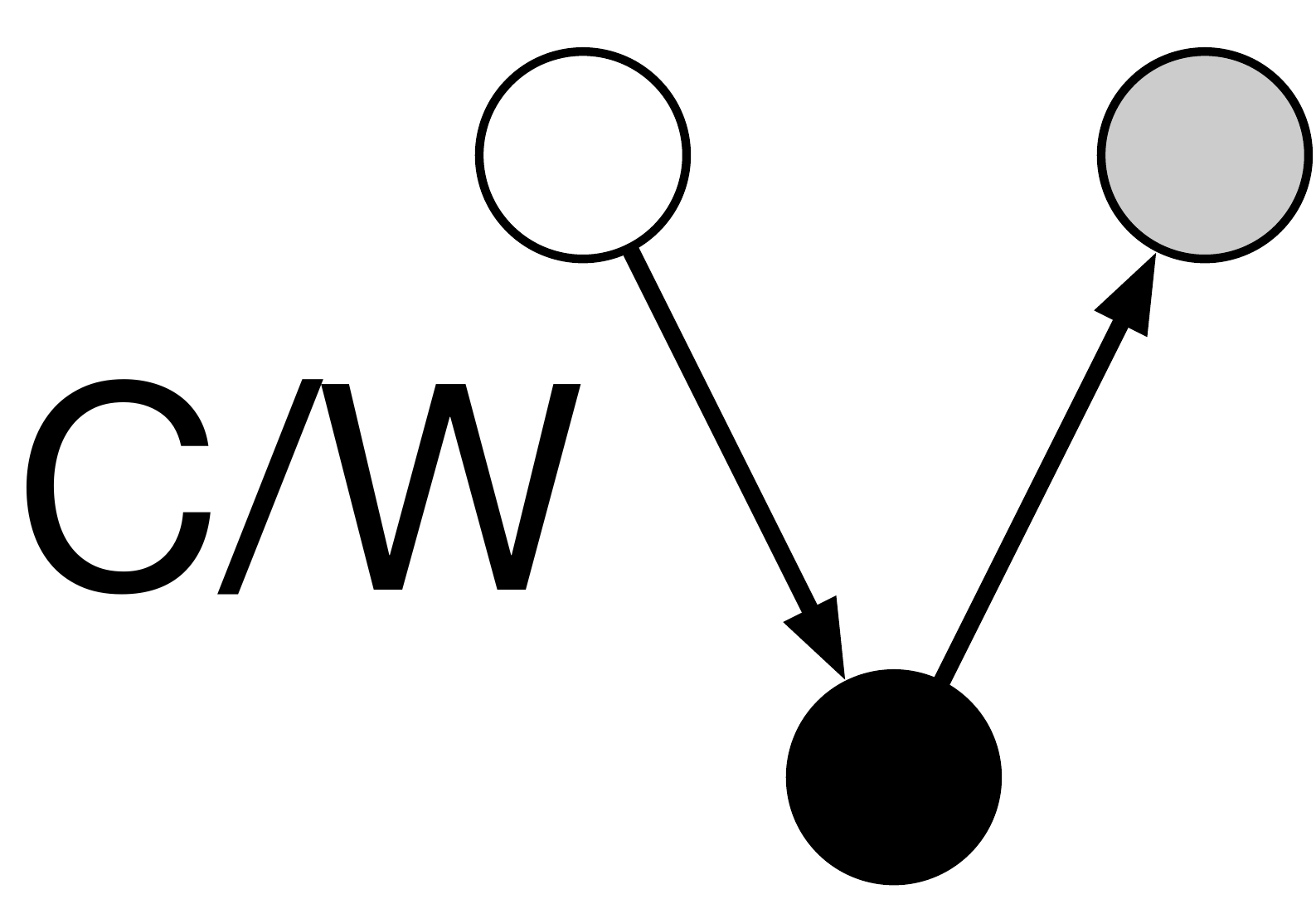}\vspace{-0.3ex}	&	46558.6	&	12313	&	68205.2	&	2594.6	&	138	&	13102.2	&	10924.6	&	238	&	50504.6	 \\ \hline
\end{tabular}
\caption{\it Size of the company for each node in the static opposition motifs. 
We show the mean, median, and standard deviation of the number of patents owned by the companies for each node in a given type of motif.  
Although the order of the events is irrelevant in static motifs, we use the same notations as those for the temporal motifs for easing the comparison with~\cref{tab:size-temporal}. In static motifs there is no repetition (therefore, no presence of R in this table), and convey and weakly-connected are the same. It should also be noted that we do not distinguish the two nodes in the ping-pong motif because of the absence of the timing information.}
\label{tab:size-static}
\end{table*}

\subsection*{Collaborations within opposition motifs}
Now we explore the interplay between patent collaborations and oppositions by looking at the 2-event opposition motifs found with $\Delta_W = $ 10 years.
For each opposition motif $M=(V', E')$ where $E'=\{(u'_1,v'_1,t'_1), (u'_2,v'_2,t'_2) \}$, we examine all the collaboration events $(a,b,t)$ such that $a, b \in V'$ and $t'_1 - 10[yr] \leq t \leq t'_2 + 10[yr]$.
Specifically, we aim to address three questions: 1) How many collaborations do we observe for a given type of opposition motif? 2) Between which two companies do we often observe collaboration? 3) When does collaboration occur and how are they related to the timing of oppositions?

\subsubsection*{Number of collaborations in opposition motifs}
To address the first question, we count the number of collaborations between each pair of companies in each 2-event opposition motif. The results are shown in~\cref{tab:count}. 
Generally, we only observe collaborations in a small fraction of the opposition motifs. 
This is because that the total number of collaborations is much smaller than the total number of oppositions (see~\cref{tab:data}).
Note that we do not observe any collaborations between two companies that have opposed each other (i.e., ping-pong motifs).
\cref{tab:count} also indicates that collaborations are substantially more frequent in out-burst and convey motifs than in in-burst and weakly-connected motifs.
We also observe that there are only three out of 475,429 out-burst motifs with 2 collaborations, and that there is no more than one collaboration in any other types of 2-event motif.
In addition, although the convey and weakly-connected motifs have the same static structure, the likelihood of collaboration is higher in the convey than weakly-connected motifs.
This may be because the two oppositions occur in a non-transitive way in the weakly-connected motifs, and therefore the two nodes with no opposition relation are more likely to be unrelated.
In contrast, the two oppositions occur in a transitive manner in convey motifs.
The scarcity of multiple collaborations is also observed in the 3-event opposition motifs; more than 91\% of them do not have any collaborations (see Supplementary Table 2). 

\begin{table*}[!t]
\centering
\begin{tabular}{|m{1 cm}||c|M{1.2cm}|M{1.2cm}|M{1.2cm}|M{1.2cm}|}
\hline
\multirow{2}{*}{Motif}    & \multirow{2}{*}{Motif count} & \multicolumn{4}{c|}{ Number of collaborations in the motif} \\
   &  & 0     & 1    & 2    & 3  \\ \hline
\includegraphics[width=1 cm]{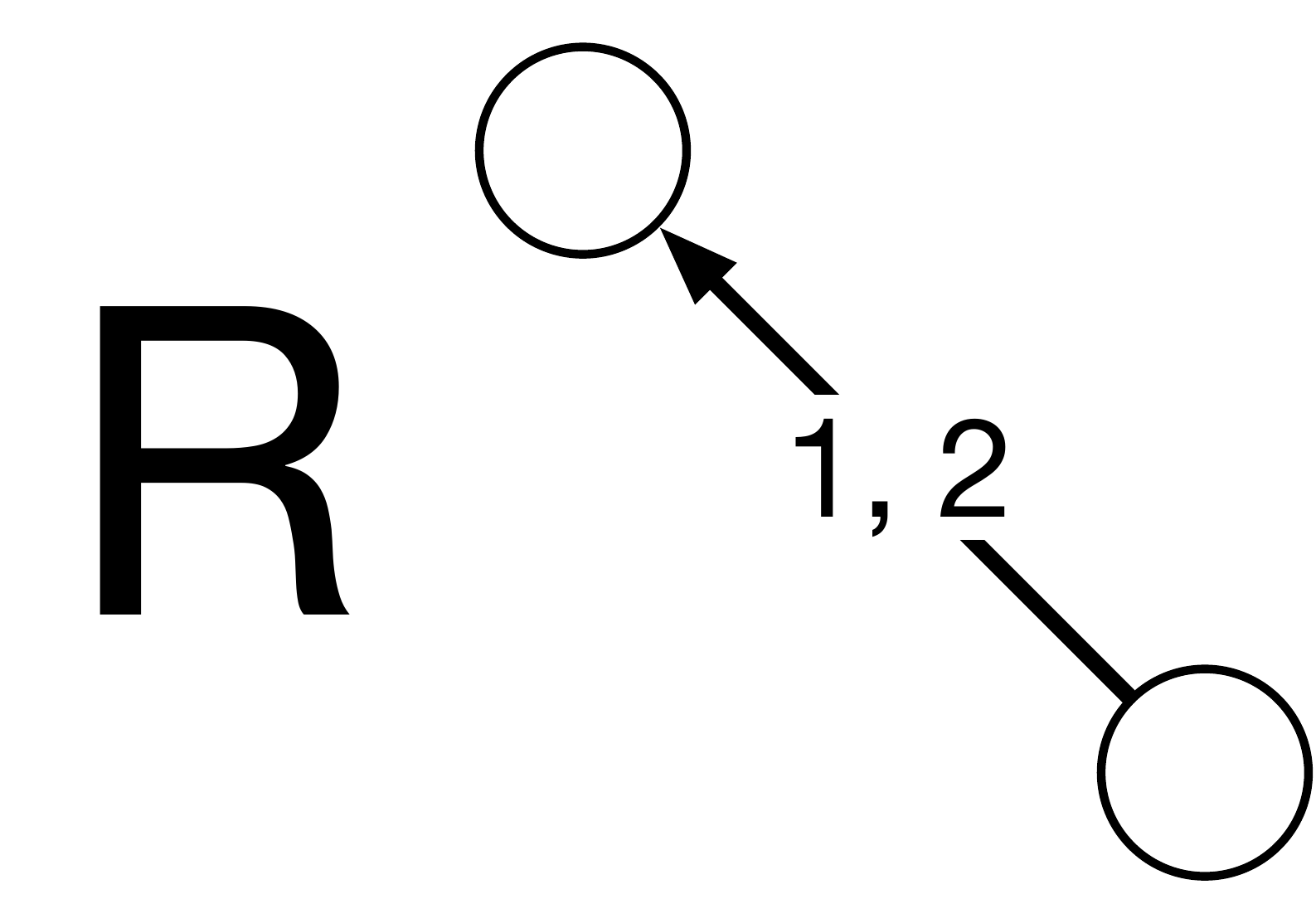}\vspace{-0.5ex} & 55,458             & 99.9\%  & 0.01\% & N/A    & N/A    \\ \hline
\includegraphics[width=1 cm]{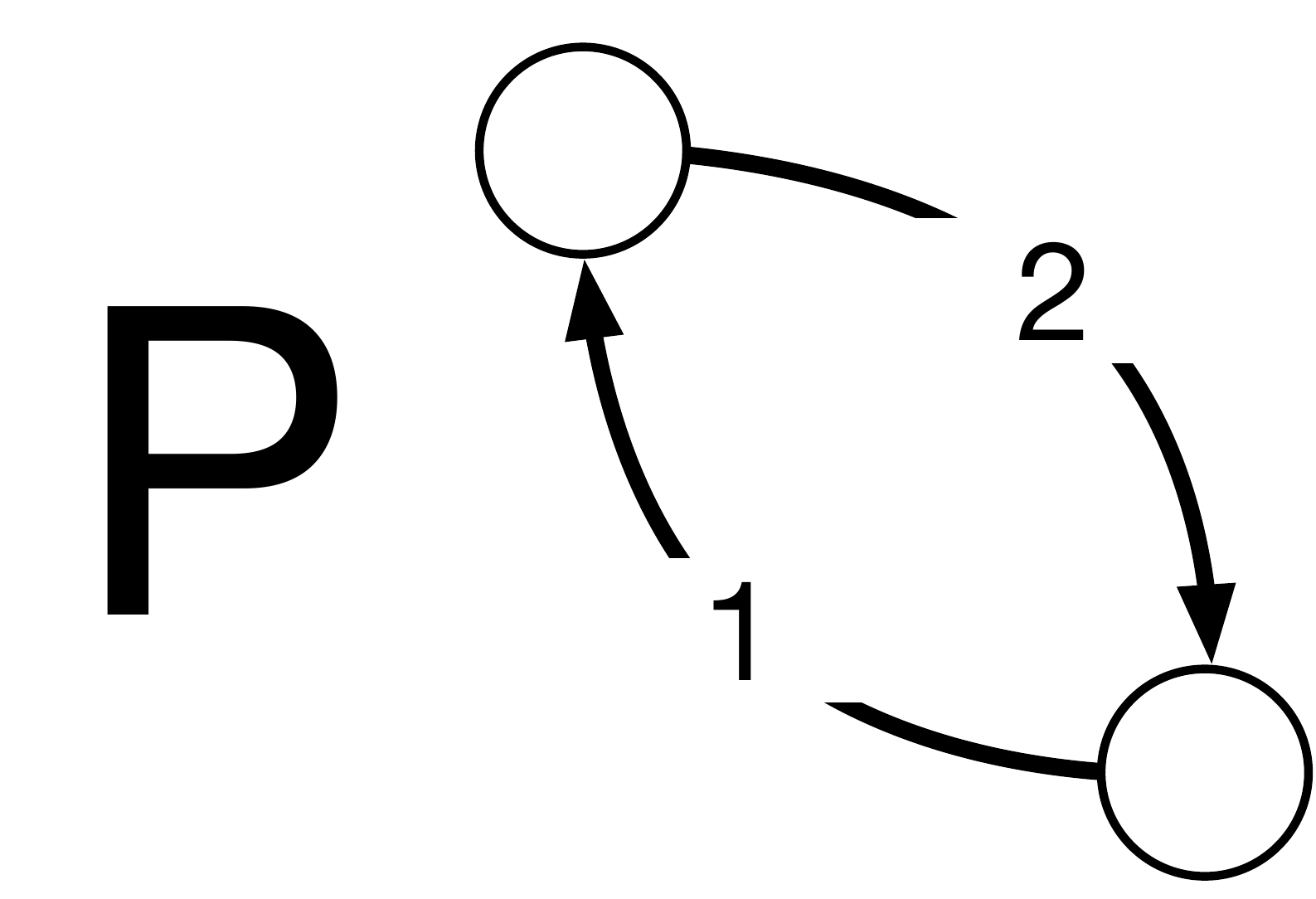}\vspace{-0.5ex} & 8,418              & 100\% & 0\% & N/A    & N/A    \\ \hline
\includegraphics[width=1 cm]{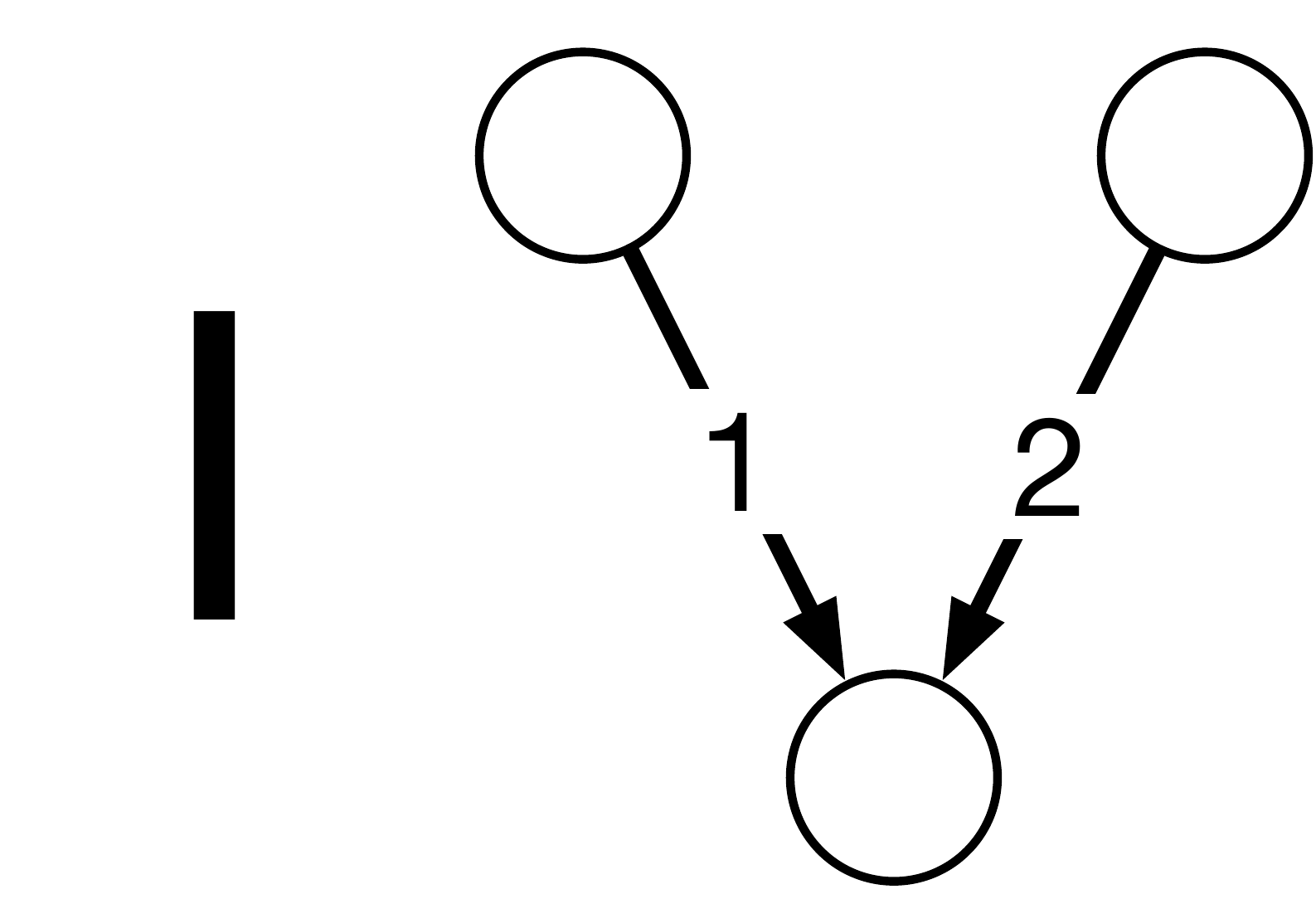}\vspace{-0.5ex} & 1,150,097           & 99.7\%  & 0.22\% & 0\% & 0\% \\ \hline
\includegraphics[width=1 cm]{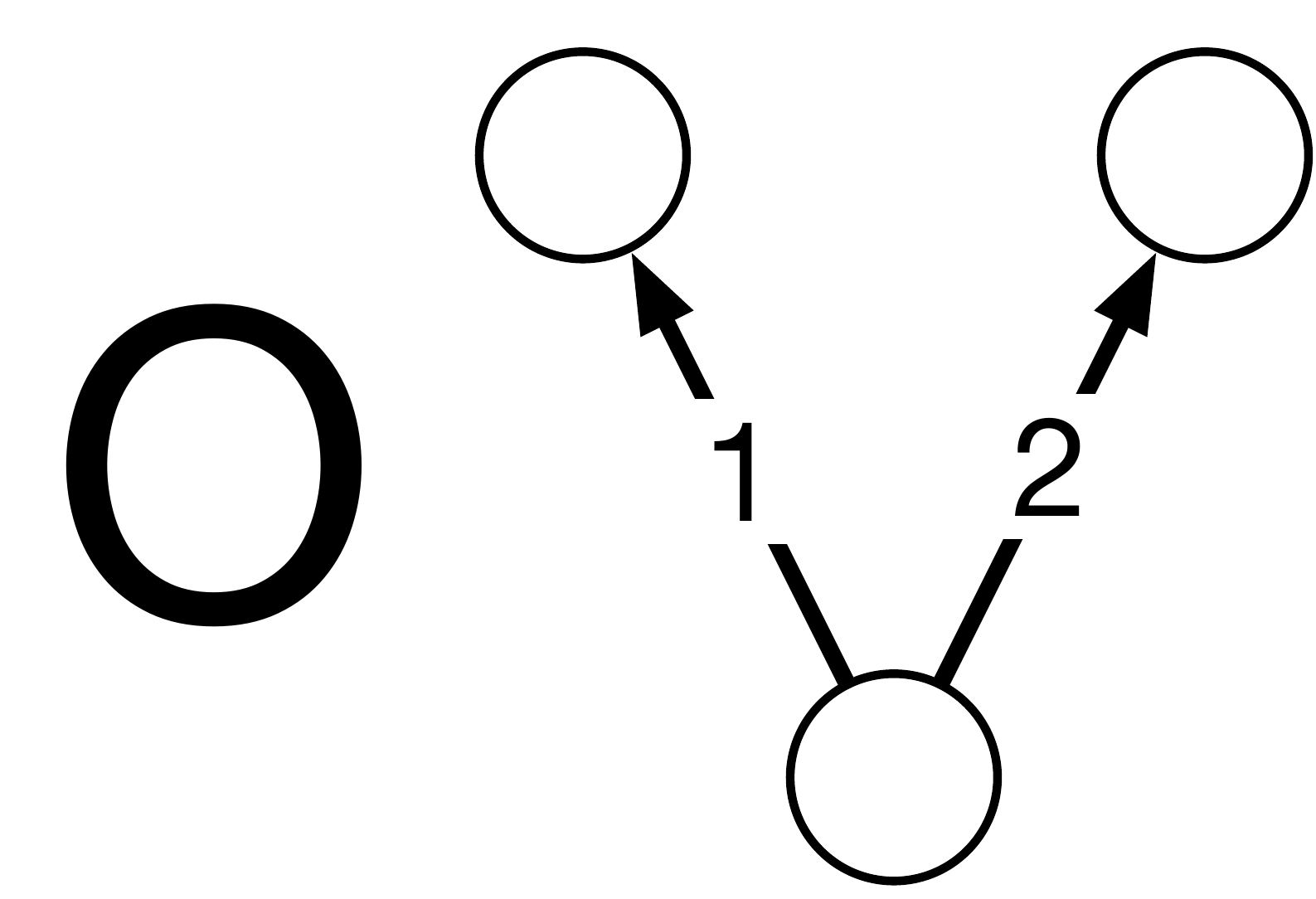}\vspace{-0.5ex} & 475,429            & 97.2\%  & 2.73\% & $<$0.01\% & 0\% \\ \hline
\includegraphics[width=1 cm]{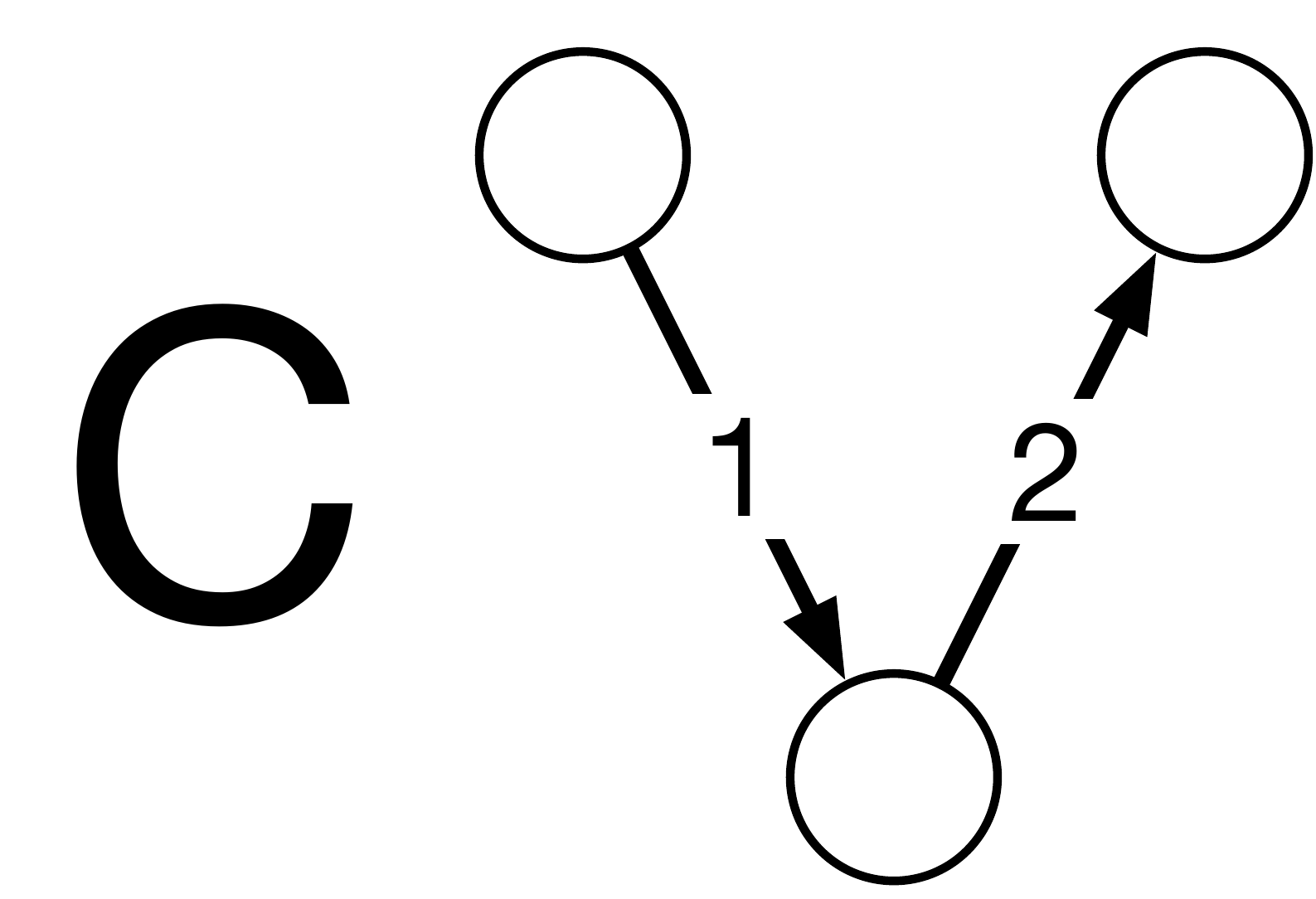}\vspace{-0.5ex} & 125,298            & 98.6\%  & 1.36\% & 0\% & 0\% \\ \hline
\includegraphics[width=1 cm]{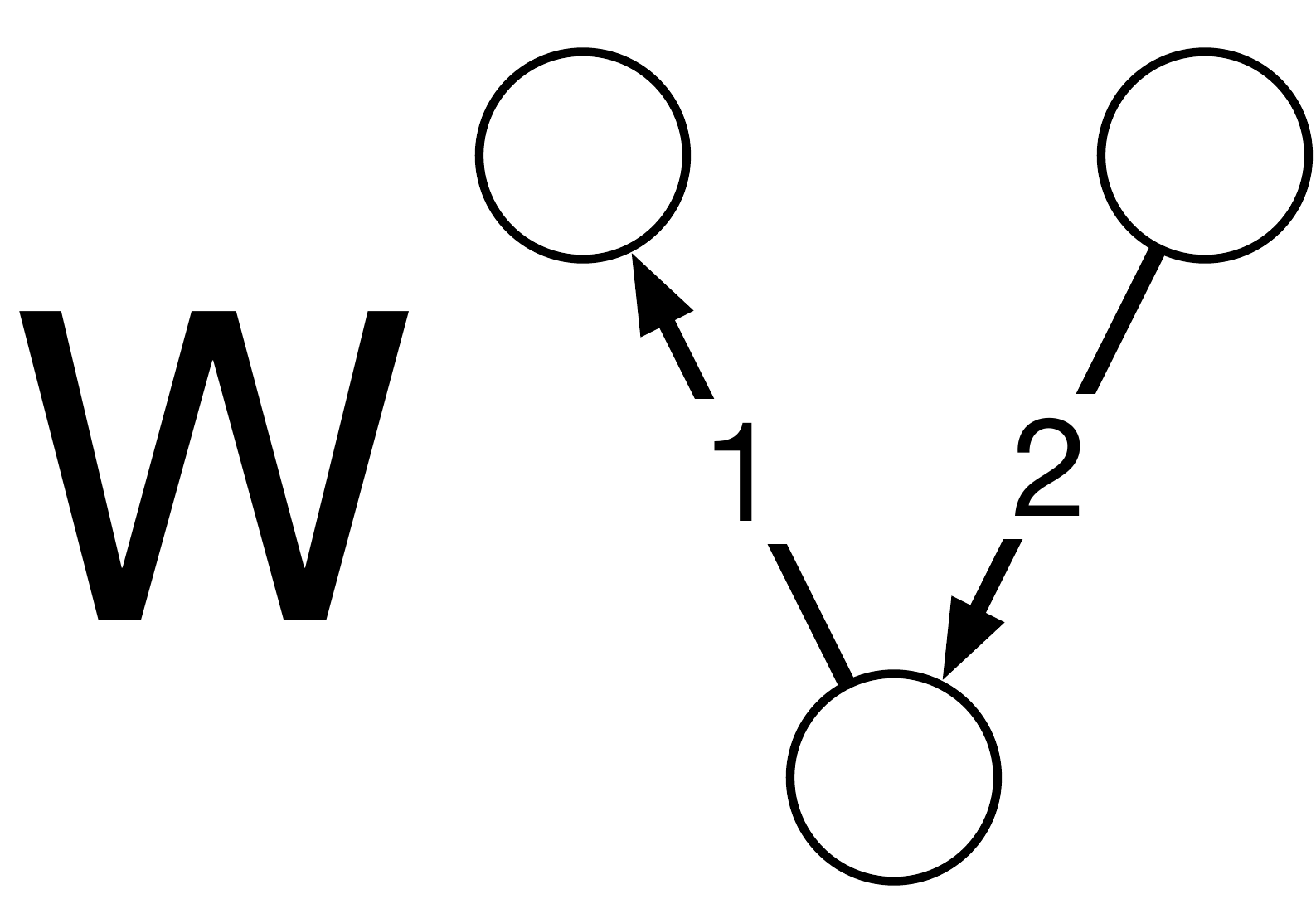}\vspace{-0.5ex} & 151,335            & 99.2\%  & 0.76\% & 0\% & 0\% \\ \hline
\end{tabular}
\caption{\it The number of collaborations observed in each 2-event opposition motif. 
Here we show the total count of each motif and the fraction of the opposition motifs with 0, 1, 2, and 3 collaborations.
}
\label{tab:count}
\end{table*}

\begin{figure*}[!b]
\centering
\includegraphics[width=0.75\linewidth]{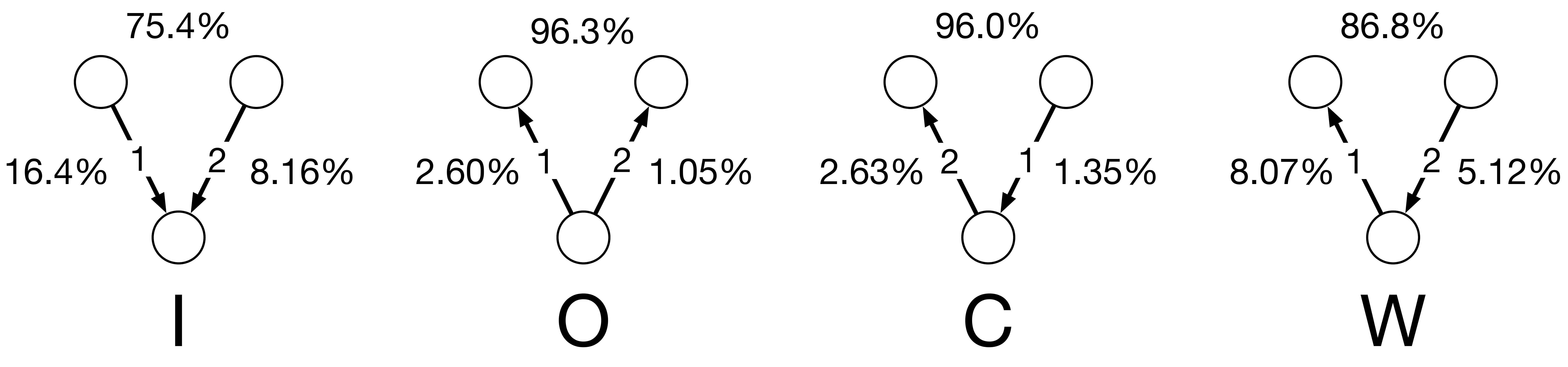}
\caption{\it Fractions of the node pairs between which the collaborations occur in 3-node 2-event opposition motifs.
}
\label{fig:where}
\end{figure*}

\subsubsection*{Companies involved in the collaborations} 

Patent opposition and collaboration are not exclusive actions. They may occur between the same pair of companies because they may collaborate in one field and compete in another field for strategic reasons. To examine this possibility, we investigate which two companies that belong to the same motif are more likely to collaborate with each other. 
We exclude the 2-node opposition motifs (i.e., repetition and ping-pong) from the following analysis because
2-node motifs only involve two companies such that which pair of companies in the motif tends to collaborate is an irrelevant question.

\cref{fig:where} shows the fractions of collaborations between each pair of companies in each type of 3-node 2-event opposition motif with at least one collaboration. 
We find that collaborations tend to occur between the companies that have no opposition relation with each other.
For example, in out-burst motifs, 96.3\% of the collaborations occur between the two nodes that have no opposition relationship.
In other words, two companies opposed by the same company are more likely to collaborate than otherwise.
Such a tendency is also present in in-burst motifs, where the two companies opposing the same company tend to collaborate with each other, but to a lesser extent (75.4\%) than in the case of out-burst motifs.
The likelihood of collaboration between the two companies with no opposition relation is larger in convey motifs than in weakly-connected motifs. 
This result may be due to the presence and absence of the transitive relationship of oppositions in the convey and weakly-connected motif, respectively.

\begin{table*}[!t]
\centering
\begin{tabular}{|m{1.1 cm}||M{1.1 cm}|M{1.1 cm}|M{1.1 cm}||M{1.1 cm}|M{1.1 cm}|M{1.1 cm}||M{1.1 cm}|M{1.1 cm}|M{1.1 cm}|}
\hline
\multirow{2}{*}{Motif}    & \multicolumn{3}{c||}{\includegraphics[width=1 cm]{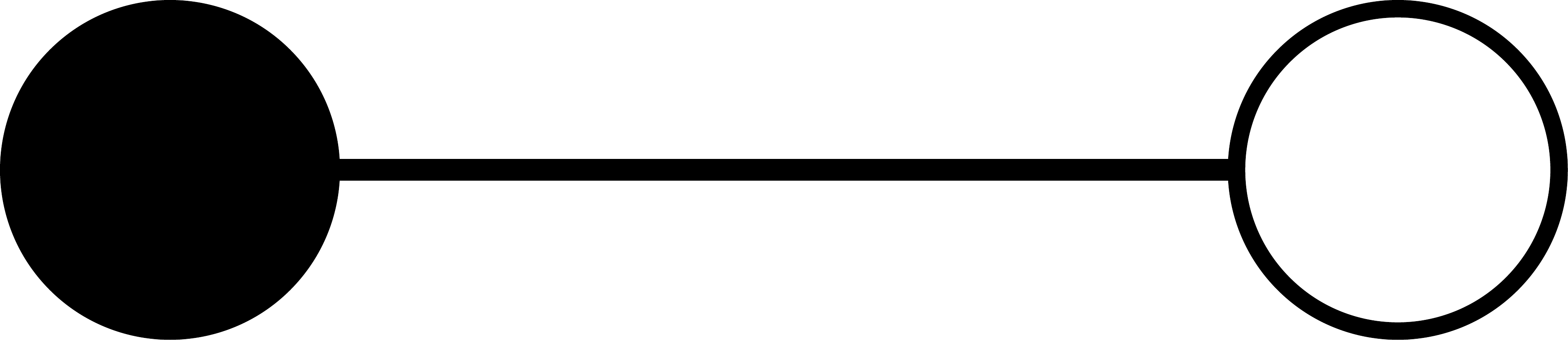}}                 & \multicolumn{3}{c||}{\includegraphics[width=1 cm]{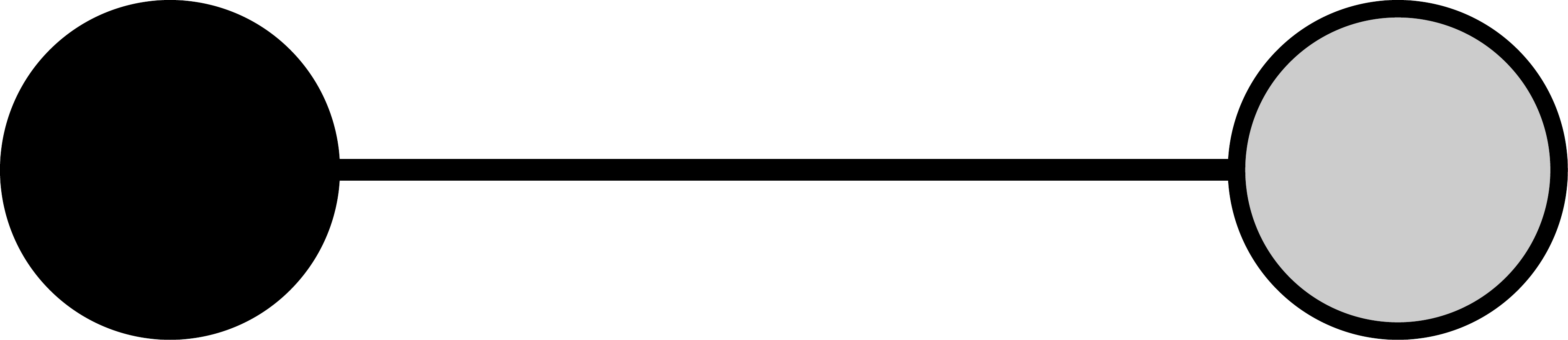}}                 & \multicolumn{3}{c|}{\includegraphics[width=1 cm]{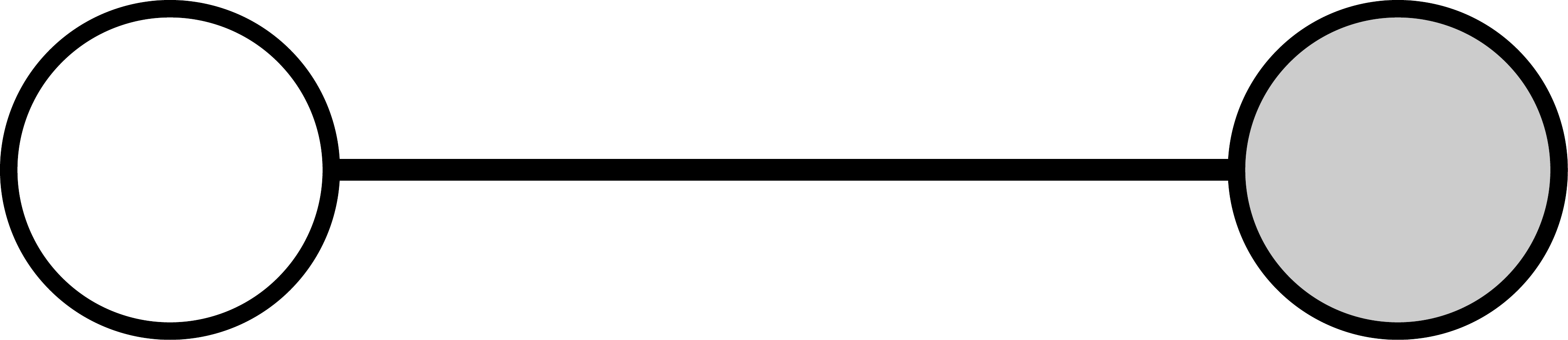}}                 \\ 
& Before & Between & After & Before & Between & After & Before & Between & After \\ \hline
\vspace{0.3ex}\includegraphics[width=1 cm]{figs-final/new-i.pdf}\vspace{-0.3ex} & 36.1\%& 12.7\%& \textcolor{red}{51.06\%}  & 17.6\%& 19.5\%& \textcolor{red}{62.8\%} & \textcolor{blue}{44.7\%}& 27.7\%& 27.5\% \\ \hline
\vspace{0.3ex}\includegraphics[width=1 cm]{figs-final/new-o.pdf}\vspace{-0.3ex} & 48.8\%& 15.9\%& \textcolor{red}{35.2\%} & 35.5\%& 16.6\%& \textcolor{red}{47.8\%} & \textcolor{blue}{41.4\%}& 34.8\%& 23.6\% \\ \hline
\vspace{0.3ex}\includegraphics[width=1 cm]{figs-final/new-c.pdf}\vspace{-0.3ex} & 13.0\%& 34.7\%& \textcolor{red}{52.1\%}& 20.0\%& 40.0\%& \textcolor{red}{40.0\%} & \textcolor{blue}{38.7\%}& 34.4\%& 26.8\%           \\ \hline
\vspace{0.3ex}\includegraphics[width=1 cm]{figs-final/new-w.pdf}\vspace{-0.3ex} & 11.8\%&17.2\%& \textcolor{red}{70.9\%} & 18.6\%& 66.1\%& \textcolor{red}{15.2\%}  & \textcolor{blue}{44.2\%}& 36.0\%& 19.7\% \\ \hline
\end{tabular}
\caption{\it Timing of collaboration events occurring on top of opposition motifs. For each type of opposition motif and each edge in the motif, we show the fraction of collaborations that occur before all oppositions, between the two oppositions, and after all oppositions. The collaborations between two nodes with an opposition relation tend to occur after all opposition events (red), whereas the collaborations between two nodes with no opposition relation tend to occur before all opposition events (blue). 
}
\label{tab:when}
\end{table*}

\subsubsection*{Timing of the collaborations}
Are collaboration events induced by the organization of opposition edges that are already existing or vice versa? To try to address this question, we investigate the temporal interrelations between opposition and collaboration events.
For a given type of 2-event opposition motif, we compute the fraction of the collaboration events occurring in three time intervals: (1) before the two opposition events, (2) between the two opposition events, and (3) after the two opposition events. We exclude the repetition and ping-pong opposition motifs from this analysis because there are a few collaboration events on top of the former and no collaboration event on top of the latter.
The results, shown in~\cref{tab:when}, indicate that, for the two companies that do not have opposition relations, the collaborations tend to happen before the two oppositions (shown in blue in~\cref{tab:when}). 
In other words, for two collaborating companies that share a common adversary, the collaborations between them tend to occur before they oppose or are opposed by a common company than vice versa.
In addition, for two companies that have both opposition and collaboration relations, while their fraction is small (see~\cref{fig:where}), the collaborations tend to occur after the two opposition events (shown in red in~\cref{tab:when}).
This result suggests that two adversarial companies are more likely to collaborate in the future than two collaborating companies oppose each other in the future.

Note that although the time intervals (i.e., before, between, and after) have the same upper bound (i.e., 10 years) as mentioned at the beginning of the section, the exact lengths of the intervals vary across the types of interval and types of motif.
For instance the mean interval length between the opposition events is around 3.9 to 4.5 years for all the motifs, which is usually smaller than the intervals before or after the two oppositions.
To calibrate for this effect, we divide each fraction in~\cref{tab:when} by the mean interval length for each case to find {\it fraction-per-year} values.
The results shown in~\cref{tab:when-normalized} support the trends found with~\cref{tab:when} albeit less strongly. In other words, collaboration between non-opposing companies tends to occur before the opposition events with a third company than after them, whereas the largest  {\it fraction-per-year} collaboration between non-opposing companies occurs between the two opposition events (see the last three columns of~\cref{tab:when-normalized}).
\cref{tab:when-normalized} also suggests that collaboration on an edge with an opposition event tends to occur after the two opposition events constituting the motif than before them (see the first six columns).

\begin{table*}[!t]
\centering
\begin{tabular}{|m{1.1 cm}||M{1.1 cm}|M{1.1 cm}|M{1.1 cm}||M{1.1 cm}|M{1.1 cm}|M{1.1 cm}||M{1.1 cm}|M{1.1 cm}|M{1.1 cm}|}
\hline
\multirow{2}{*}{Motif}    & \multicolumn{3}{c||}{\includegraphics[width=1 cm]{figs-final/e-01-new.pdf}}                 & \multicolumn{3}{c||}{\includegraphics[width=1 cm]{figs-final/e-02-new.pdf}}                 & \multicolumn{3}{c|}{\includegraphics[width=1 cm]{figs-final/e-12-new.pdf}}                 \\ 
& Before & Between & After & Before & Between & After & Before & Between & After \\ \hline
\vspace{0.3ex}\includegraphics[width=1 cm]{figs-final/new-i.pdf}	&	4.32\%&	2.96\%&	{5.38\%}&	2.10\%&	4.54\%&	{6.61\%} &	5.34\%&	{6.45\%} &	2.90\%	\\ \hline
\vspace{0.3ex}\includegraphics[width=1 cm]{figs-final/new-o.pdf}	&	{5.23\%} &	4.06\%&	4.89\%&	3.80\%&	4.24\%&	{6.64\%} &	4.44\%&	{8.89\%}&	3.28\%	\\ \hline
\vspace{0.3ex}\includegraphics[width=1 cm]{figs-final/new-c.pdf}	&	1.35\%&	{8.34\%}&	7.63\%&	2.08\%& {9.62\%} &	2.93\%&	4.02\%&	{8.27\%} &	3.93\%	\\ \hline
\vspace{0.3ex}\includegraphics[width=1 cm]{figs-final/new-w.pdf}	&	1.22\%&	3.76\%&	{9.99\%} &	1.92\%& {14.46\%} &	2.14\%&	4.55\%&	{7.88\%} &	2.78\%	\\ \hline
\end{tabular}
\caption{\it Normalized timing of collaboration events occurring on top of opposition motifs. For each type of opposition motif and each edge in the motif, we show the fraction of collaborations per year that occur before all oppositions, between the two oppositions, and after all oppositions. For each interval, the fraction is normalized by the length of the interval.
}
\label{tab:when-normalized}
\vspace{-3ex}
\end{table*}

\section*{Discussion}

We studied the adversarial and collaborative relations between the patent holder companies. 
We first analyzed the frequency of patent oppositions and showed that (1) the oppositions tend to be more repetitive than reciprocal; (2) different timing thresholds between the oppositions present a consistent behavior in the ranking of 2-event and 3-event motif types; and (3) oppositions exhibit coherent results with the structural balance theory.
We also measured the statistical significance of opposition motif frequencies using different null models and showed that the motifs containing repetitions and in-bursts tend to be abundant, whereas the ping-pong and triangle motifs are scarce.
Our analysis of the company sizes in the opposition motifs showed that (1) the opposer companies tend to be larger than the opposed companies in most 2-event motifs; and (2) large companies are likely to be engaged in oppositions with multiple companies.
In our analysis on the interplay between collaborations and oppositions, we found that (1) collaborations tend to occur between the companies that have no opposition relation with each other; (2) for two collaborating companies that share a common adversary, the collaborations between them tend to occur before they oppose or are opposed by a common company than vice versa; and (3) two adversarial companies are more likely to collaborate in the future than two collaborating companies oppose each other in the future.

One limitation of our work is that we considered only the 2-event and 3-event motifs with up to 3 nodes. 
Extending the size of the motifs may reveal new insights regarding the correlation and interplay between collaborations and oppositions.
Moreover, the nodes in the motifs can be enriched with the country and sector information of the companies, and the events in the motifs can be enhanced by the features of the patent and opposition texts.
Such information can potentially lead to a more holistic assessment of oppositions and collaborations.

Another potential opportunity for future research is designing new null models for temporal networks.
In our analysis, all the null models yielded largely positive $Z$ scores---the counts in the empirical networks are always larger than those in the null models.
Choosing a proper null model is already challenging for static networks~\cite{artzy2004comment} and the problem is even more difficult for temporal networks due to the intertwined temporal and structural complexity~\cite{K11, gauvin2018randomized}.
Deploying and assessing null models for a more rigorous measurement of the statistical significance of temporal motifs also warrants future work.

\bibliography{paper,afosr}

\section*{Acknowledgements}
P.L., N.M., and A.E.S. acknowledges the support by JP Morgan Chase and Company Faculty Research Award.

\section*{Author contributions statement}
N.M., T.K., and A.E.S. conceived the research; N.M., T.K., and A.E.S designed the research; T.K. collected the data; P.L. and T.K. analyzed the data; P.L., N.M., T.K., and  A.E.S. discussed the results and wrote the paper.

\section*{Competing interests}
The authors declare no competing interests

\end{document}